\documentclass[aps,twocolumn]{revtex4}
\usepackage{graphicx}
\usepackage{amsmath}   
\usepackage{epstopdf}
\def\bomega{\mbox{\boldmath $\omega$}}
\def\bbomega{\mbox{\boldmath $\Omega$}}
\def\bsigma{\mbox{\boldmath $\sigma$}}

\begin{document}
\title{Effect of 1/f noise on the dissipative dynamics of an LC-shunted qubit}
\author{ F. T. Vasko}
\email{ftvasko@gmail.com}
\affiliation{QK Applications, San Francisco, CA 94033, USA }
\date{\today}

\begin{abstract}
We consider dissipative dynamics of a flux qubit caused by 1/f noises, which act both on the shunting LC-contour and on the SQUID loop. These classical Gaussian noises modulate of the level splitting and of the tunnel coupling, respectively, and they are partially correlated. The transient evolution of qubit has been studied for the regimes: (a) the interwell incoherent tunneling, (b) the relaxation of interlevel population, and (c) the decoherence of the off-diagonal part of a density matrix. For all regimes, the relaxation rates and the frequency renormalization [for the case (c)] are analyzed versus the parameters of qubit and  couplings to the noises applied. The fluctuation effects give a dominant contribution at tails of relaxation, so that the averaged dissipative dynamics is not valid there. The results obtained open a way for verification of the parameters of qubit-noise interaction and for minimization of coupling between qubit and environment. Under typical level of noises, the results are comparable to the recent experimental data on the population relaxation and on the incoherent interwell tunneling.
\end{abstract}

\maketitle

\section{Introduction}
During last decade, an essential progress was made towards the implementation of the quantum information protocols, see \cite{1,2,3,4,5} and references therein. An essential part of these results are based on different types of the superconducting flux qubits. Whereas the dynamic properties of the noiseless qubits are effectively analyzed with the use of the lumped-element approach, \cite{6,7,8} both the mechanisms of the qubit-environment interaction and the dissipative dynamics of qubits are not investigated completely. Heretofore, a partial characterization of qubits is performed with the use of the spectroscopy in GHz region, for different regimes of the high frequency response and of the readout \cite{9,10,11,12}. Beside of this, the incoherent resonant tunneling, both between the anticrossing levels (the Landau-Zener transitions \cite{13}) and between the steady-state levels \cite{14}, or the low-frequency (sub-kHz) measurements  \cite{14a} are employed for the study. The experimental data suggest that the qubit-environment interaction is caused by the low  frequency classical noise (described by the 1/f spectral function)  and the high-frequency bosons (described by the quasi-ohmic spectral function), see analysis in \cite{10,15}. Recent studies of these processes were directed on an improvement of the fidelity of the quantum hardware. But any consideration based on a simplified model of the spin-noise interaction (e.g., see \cite{16,17} and references therein) is not enough for applications to multi-qubit hardware because such results cannot be associated to a set of parameters describing a real  device.
%f1
\begin{figure}[ht]
\begin{center}
\includegraphics[scale=0.22]{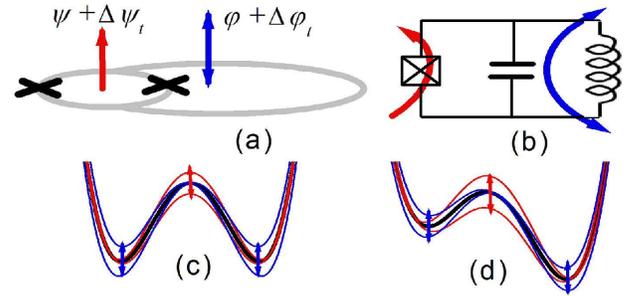}
\end{center}
\addvspace{-0.5 cm}
\caption{ (a) Sketch of qubit formed by the Josephson junction loop shunted by the transmission line. Tilt and control fluxes, $\varphi$ and $\psi$, with random contributions, $\Delta\varphi_t$ and $\Delta\psi_t$, are shown by the blue and red arrows, respectively. Regimes of relaxation are different for parallel or anti-parallel directions of fluxes $\varphi$ and $\psi$. (b) Model circuit involving the effective Josephson junction shunted by the  effective LC-contour under external fluxes $\varphi +\Delta\varphi_t$ and $\psi +\Delta\psi_t$ (blue and red). (c) Modulation of the symmetric potential energy (thick curve) via tilt ($tl$-) and control ($j$-) channels (red and green, respectively) resulting in variations of level splitting and barrier height, respectively. (d) The same as in panel (c) for the tilted potential, $\varphi\neq 0$. } 
\end{figure} 

Thus, it is timely to study quantum hardware based on the lumped element approach involving an effective circuit with a detailed description of the qubit-environment interaction. Under such a consideration, the twofold aim should be achieved: (i) to introduce and to justify {\it a realistic model} of the qubit-noise interaction, which is beyond of the simplified models, and (ii) to examine the dissipative dynamics of qubit in order to perform {\it a complete characterization} of this interaction through the analysis of transient processes.  Such an analysis opens a way to optimize relaxation parameters and to enhance a fidelity of the quantum hardware.

In this paper, we perform a complete analysis of the dissipative dynamics for the qubit formed by the SQUID loop shunted by the transmission line (so called fluxmon \cite{15}), see sketch in Fig. 1(a). Characteristics of such a qubit are governed by the tilt and control fluxes, $\varphi$ and $\psi$, applied through the transmission line and the SQUID loop, respectively. We restrict ourselves by the low-frequency region when interactions with 1/f noises are described by the classical random fluxes, $\Delta\varphi_t$ and $\Delta\psi_t$, which are added to the tilt and control fluxes. The calculations of transient evolution are performed for the model of the qubit used the effective Josephson junction shunted by the effective LC-oscillator, see circuit in Fig. 1(b). We employ the two-level approach taking into account {\it a softness of the double-well potential}, in contrast to the solid-state models \cite{18} based on a fixed tunnel-coupling (due to the rigid barrier), and a variable level-splitting. The shape of potential is controlled by an interplay of the inductive and Josephson energies as it is shown in Figs. 1(c) and 1(d) for the symmetric ($\varphi =0$) and tilt ($\varphi\neq 0$) cases; the thin color curves draw a noise-induced modulation of potential via a shift of minima and a modulation of  barrier height. Averaged dynamics of the qubit is determined by the correlation functions $\langle\Delta\varphi_t\Delta\varphi_t\rangle$, $\langle\Delta\psi_t\Delta\psi_t\rangle$, and $\langle\Delta\varphi_t\Delta\psi_t\rangle$, which describe noises excited in the transmission line and in the SQUID loop ($tl$- and $j$-channels) as well as {\it correlations between these noises} ($c$-channel) \cite{19}. These correlations may caused by a common external sources or an inductive coupling between $tl$- and $j$-channels. We use the 1/f spectral functions $\alpha_k /\omega$ with $k=(tl,j,c)$, so that the dissipative dynamics is determined by the parameters of qubit, the dimensionless noise strengths $\alpha_k$, and the control fluxes, $\varphi$ and $\psi$.

The two lowest electromagnetic modes of the flux qubit are described by the isospin variable and the exact dynamics of qubit is governed by the Bloch equations with a time-dependent rotation frequency. We analyze the transient evolution of the density matrix at $t>0$ for the regimes: (a) resonant tunneling between the left and right wells, when the noise-induced broadening exceeds the interlevel energy, (b) relaxation of the interlevel population, and (c) decoherence of the off-diagonal part of a density matrix. The cases (b) and (c) correspond to the weakly-coupled levels, when the gap frequency exceeds the noise-induced broadening of levels \cite{20}. The Bloch equations for the case (a) is solved with the use of the symmetric qubit frame when the eigenstate problem is solved numerically at $\varphi =0$. For the cases (b) and (c) we use the tilt ($\varphi$-dependent) frame, when the basis functions correspond the lower and upper levels. For all cases, it is convenient to transform the Bloch system into the single integro-differential equations with the different kernels. Within  the weak-fluctuation approach, the averaged responses are written through the Laplace transforms of the averaged kernels and the parameters of relaxation are obtained in the explicit forms. Straightforward calculations of the fluctuating corrections for the cases (a)-(c) give {\it the limitations at tails of relaxation} for the averaged approach outlined above.

Based on this model, we have performed a complete analysis of the incoherent tunneling rate, the population relaxation rate, and the decoherence rate, $\nu_{T}$, $\nu_{1}$, and $\nu_{2}$ (i.e. of the inverse times of the different relaxation processes), as well as the noise-induced renormalization of the gap frequency. The main new results are summarized as follows: 
\begin{itemize}
\item[(1)]\vspace{-0.25 cm} The averaged dissipative dynamics is governed by the non-local in time equations [Eqs. (11), (21), and (27) below] during the initial stage of relaxation $\nu_{T,1,2}t\leq 1$, while the exponential decay takes place if $\nu_{T,1,2}t\geq 1$. Peaks of the relaxation rates between wells or between levels, $\nu_T$ or $\nu_1$ versus tilt $\varphi$, are due to the noise in $tl$-channel while the dip of decoherence rate $\nu_2$ near $\varphi =0$ is determined by an interplay of the noises in $tl$- and $j$-channels. Weak asymmetry of $\nu_{T,1,2}$ around $\varphi =0$ is determined by the $c$-channel, i.e. by correlations between the $tl$- and $j$-noises. 
\item[(2)]\vspace{-0.25 cm} The half-width of the Gaussian peak of incoherent tunneling is $\propto \sqrt {\alpha_{tl}}$ while the amplitude is $\propto 1/\sqrt{\alpha_{tl}}$. Weak asymmetry of this peak is $\propto\alpha_c /\alpha_{tl}$ and the direction of shift (along $\varphi >0$ or $<0$) is determined by the orientation of fluxes $\varphi$ and $\psi$ [parallel or anti-parallel, see Fig. 1(a)]. In addition to the mechanism considered in Refs. 14 and 22, the classical 1/f noise considered here provides an essential contribution to the experimental data reported in Ref. 16.
\item[(3)]\vspace{-0.25 cm}  The population relaxation rate is $\propto\alpha_{tl}$ with tails of peak suppressed as $\varphi^3$ and the parameters of peak are comparable to the recent experimental data \cite{15}. Due to softness of the barrier, there is no the dependency $\nu_1\propto\omega_{10}^{-1}$ for the 1/f noise under consideration. This disagreement with \cite{15} may be compensated by the frequency dispersion of noise caused by the size effect in transmission line \cite{22}. 
\item[(4)]\vspace{-0.25 cm} There is {\it a dip of decoherence} : the rate $\nu_2$ and the renormalization of the gap frequency are increased sharply around $\varphi =0$ and are saturated with increasing of $\varphi$. There is the two-mode oscillating decoherence regime where the depth and asymmetry of the dip are determined by the ratios $\alpha_j /\alpha_{tl}$ and $\alpha_c /\alpha_{tl}$, respectively.
\item[(5)]\vspace{-0.3 cm} The averaged descriptions fail at tails of transient relaxation because of {\it the rare fluctuations effect}. The mean-square-fluctuation of the interwell tunneling increases $\propto t^2$ and, for typical parameters, the average description of incoherent tunneling is valid at $\nu_T t<3$. There are time-independent levels of fluctuations for the interlevel population and the decoherence [cases (b) and (c)] with the logarithmically enhanced contribution of rare fluctuations. Under tilt, $\varphi\neq 0$, the fluctuation effects are suppressed for all regimes. 
\end{itemize}\vspace{-0.25 cm}

The paper is organized as follows. The model of the LC-shunted qubit, which interacts with partially correlated noises in the tilt and control channels, is presented in Sec. II. The resonant incoherent tunneling between wells is considered in Sect. III. The population relaxation rate and the  decoherence process are analyzed in Sect. IV for the weak-coupling regime. For all cases, the levels of fluctuations are considered in Sect. V. The concluding remarks, the list of assumptions, and the discussion of current experimental context are given in the last section. The averaged kernels are calculated in Appendix taking into account $tl$-, $j$-, and $c$-channels.

%%%%%%%%%%%%%%%%%%%%%%%%%%%%%%%%%%%%%%%
\section{Qubit-noise interaction }
We start with the description of the flux qubit, formed by SQUID shunted by LC-contour which are interacted with 1/f noises. Quantum mechanics of such a qubit in the $\phi$-representation is described by the Hamiltonian:
%1
\begin{equation}
\hat H =4E_C{\hat q^2} +E_L (\phi -\varphi )^2 /2+{E_J}\cos (\psi /2)\cos\phi ~.
\end{equation}
Here $\phi$ and $\hat q=-id/d\phi$ are the dimensionless flux and the charge operator, while $\varphi$ and $\psi$ are the external tilt and control fluxes penetrated through the LC-contour and the SQUID loop, respectively. We use the dimensionless variable $\phi$ as well as the control fluxes $\varphi$ and $\psi$ in units $\Phi_0 /2\pi$ where $\Phi_0 =\pi\hbar /|e|$ is the flux quantum. Eq. (1) involves the capacitance energy $\propto E_C =e^2 /2C$, the inductive energy $\propto E_L =(\Phi_0 /2\pi )^2 /L$, and the effective Josephson energy $\propto E_J =I_c\Phi_0 /2\pi$ which describes the SQUID loop with the critical current $I_c$ and can be varied via the factor $\cos (\psi /2)$. \cite{23} If energy scale is fixed in units $E_L$, the Hamiltonian (1) is dependent on the external fluxes $\varphi$ and $\psi$, as well as on the parameters $4E_C /E_L\ll 1$, and $\beta_\psi =E_J\cos (\psi /2) /E_L\sim 1$. Noise contributions are introduced under the replacements of external factors  in Eq. (1) by $\varphi\to \varphi +\Delta\varphi_t$ and $\psi\to\psi +\Delta\psi_t$. The Hamiltonian of qubit-noise interaction, 
%2
\begin{equation}
\hat H_{\rm int}=-\Delta\varphi_t E_L (\phi -\varphi )-\Delta\psi_t E_J\sin (\psi /2)(\cos\phi )/2~,   
\end{equation}
is written within the linear approach with respect to dimensionless random contributions, $\Delta \varphi_t$ and $\Delta\psi_t$. Notice, that energy spectrum determined by the noiseless Hamiltonian (1) is not changed under the separate replacements $\varphi\to -\varphi$ (with $\phi\to -\phi$) or $\psi\to -\psi$. In contrast,  $\hat H_{\rm int}$ is is not changed if  only $\varphi\to -\varphi$ and $\psi\to -\psi$, i.e. the relaxation processes are different for the parallel or antiparallel directions of the external fluxes.

Under averaging over random contributions $\Delta\varphi_t$ and $\Delta\psi_t$, we consider the Gaussian random processes taking into account a partial correlation between noises in the $tl$- and $j$-channels. Introducing the 1/f spectral function of the $k$th channel as $\alpha_k /\omega$, one obtains 
%3
\begin{eqnarray}
\left| \!{\begin{array}{*{20}c}
{\left\langle {\Delta \varphi _t \Delta \varphi _{t'} } \right\rangle }  \\
   {\left\langle {\Delta \psi_t \Delta \psi_{t'} } \right\rangle }  \\
   {\left\langle {\Delta \psi_t \Delta \varphi _{t'} } \right\rangle }  \end{array}}\! \right| = \left| {\begin{array}{*{20}c} {\alpha _{tl} }  \\ {\alpha _{j} }  \\ {\alpha _{c} } \end{array}} \right| \int\limits_{\omega_m}^{\omega_M}\!\frac{d\omega}{\pi\omega}\cos\omega\Delta t \! \equiv\! \alpha_k w_{\omega _m \Delta t}  \nonumber	\\
\approx\frac{\alpha_k}{\pi} \left\{\ln\frac{1}{\omega_m \Delta t}-\gamma + o\left[ {(\omega _m \Delta t)^2 } \right] \right\} ~, ~~~~~ 
\end{eqnarray}
where $\alpha_k$ determines a coupling strength in $k$th channel  ($k=tl, ~j, ~c$), $\Delta t=t-t'$, and we use the cut-off frequencies $\omega_{M,m}$ which are supposed to be the same in all channels if $\omega_M^{-1}\ll\Delta t\ll\omega_m^{-1}$. The explicit expression of the correlator $w_{\omega_m|\Delta t|}$ is approximately written through the integral cosine which has the logarithmic asymptotic at $\omega_m\Delta t\ll 1$ with the Euler's constant, $\gamma\approx 0.577$ \cite{24}. Thus, qubit-environment interaction is characterized by three constants, $\alpha_k$, and the lower cut-off frequency $\omega_m$.  

Below we use the basis determined by the symmetric ($\varphi =0$) eigenstate problem $\hat H_{\varphi =0,\beta}\left| n\right\rangle =\varepsilon_{n}\left| n\right\rangle$, where $\hat H_{\varphi ,\beta}=\hat H -\hat H_{\rm int}$. In {\it the symmetric qubit frame}, the solutions $\left| n\right\rangle$ and $\varepsilon_{n}$ are parametrically dependent on the control flux through $\beta_\psi$ and $\varphi =0$. The density matrix takes form $\hat\rho_t =\sum \nolimits_{n_1 n_2}{\rho_{n_1 n_2}}{\left| {n_1 } \right\rangle }\left\langle {n_2} \right|$, where $\rho_{n_1 n_2}$ is governed by the standard equation $i\hbar\partial\rho_{n_1 n_2} /\partial t =[\hat H_t ,\hat\rho_t ]_{n_1 n_2}$ with the effective Hamiltonian, 
%4
\begin{equation}
H_{nn'}=\varepsilon_n\delta _{nn'}-E_L\varphi (\phi )_{nn'}+\left(\hat H_{\rm int}\right)_{nn'}  ~,     
\end{equation} 
which is written through matrix elements $(\cdots )_{nn'}\equiv\left\langle n \right|\cdots\left| {n'} \right\rangle$. Note, that $\varepsilon_n$ is determined for the symmetric barrier, while the tilt effect is described here by the off-diagonal matrix element $\propto (\phi )_{nn'}$. We employ below the two-level approach and consider only the states $\left| 0\right\rangle$ and $\left| 1\right\rangle$. Introducing the time-independent gap frequency $\omega_{10}=(\varepsilon_1 -\varepsilon_0 )/\hbar$ and the Pauli matrices $\hat\bsigma$, which are written here with respect to the basis formed by the $|0\rangle$ and $|1\rangle$ states, we arrive to the $2\times 2$ matrix Hamiltonian  
%5
\begin{eqnarray}
\frac{\hat H_t}{\hbar}\! =\!\frac{(\bomega_t\cdot\hat{\bsigma})}{2} , ~~  \bomega_t\! =\! \left[ -2\widetilde\omega (\varphi +\Delta \varphi_t ), 0, \omega_{10} -\omega_j \Delta\psi_t \right] , \nonumber   \\
\widetilde \omega \!  =\! \frac{E_L (\phi )_{10}}{\hbar} ~, ~~~~ \omega_j \! =\! \frac{E_J}{2\hbar} (\cos \phi )_{11,00}\sin\frac{\psi}{2} ~.  ~~~~ 
\end{eqnarray}
Here the frequencies $\widetilde\omega$ and $\omega_j$ determine the additional tunneling mixing of the $|0\rangle$ and $|1\rangle$ states due to the total tilt flux, $\varphi +\Delta \varphi_t$, and modulation of $\omega_{10}$ by the noise due to barrier fluctuations, respectively. We also omit the contributions $\propto$ const and take into account that the wave functions $\langle \phi | 0\rangle$ and $\langle\phi | 1\rangle$ are symmetric and antisymmetric with respect to $\phi \leftrightarrow  - \phi$. The matrix elements $(\phi )_{10}$ and $(\cos\phi )_{11,00}\equiv (\cos\phi )_{11}-(\cos\phi )_{00}$ are dependent on $\beta_\psi$ and $\omega_j$ is written through $\sin (\psi /2)$.

The projection operators on the left/right ($l/r$) wells in $\phi$-representation with the Hamiltonian (1) are $\hat P_{l/r}  = \theta ( -\! /\!+\phi )$, so that $\hat P_l +\hat P_r =1$ and  $\hat P_l\cdot\hat P_r =0$. Within the two-level approach describing by the Hamiltonian (5), these operators are given by 
%6
\begin{equation}
\hat P_{l/r}  = \left| \begin{array}{*{20}c}
  {\left\langle {1|1} \right\rangle } & {\left\langle {1|0} \right\rangle }   \\  {\left\langle {0|1} \right\rangle }  & {\left\langle {0|0} \right\rangle }  \end{array} \right|_{\phi <0 /\phi >0 }\approx \frac{1-\! /\! + \hat\sigma_x}{2} ~,
\end{equation} 
where $\left\langle {1|1} \right\rangle_{\phi <0 /\phi >0}=\left\langle {0|0} \right\rangle_{\phi <0 /\phi >0}=1/2$ due to the symmetry reason. Numerical estimates of the off-diagonal matrix elements with the eigenstates calculated below (see Fig. 2 below) give $0.45<|\left\langle{1|0}\right\rangle_{\phi <0 /\phi >0}|<0.5$ if $\beta >1.1$ and $|\hat P_l\cdot\hat P_r |<0.05$ instead of the exact zero. This inconsistency is due to we drop the upper states, $n>1$, and below in Sect. III we use the right-hand expression of Eq. (6) which corresponds the weak-tunnel-coupling case. 

If tunneling mixing increases, it is convenient to use  {\it the tilt qubit frame}. After rotation of $\hat H_t$ around 0Y-axis Eqs. (5) are transformed into 
%7
\begin{eqnarray}
\frac{\hat{\cal H}_t}{\hbar} =\frac{(\bbomega_t\cdot\hat{\bsigma})}{2}, ~~ \bbomega_t =\left[ -\frac{\Delta\varphi_t 2\widetilde\omega \omega _{10} +\Delta\psi_t\omega _j \widetilde\omega \varphi}{\Omega_\varphi} \right. , \nonumber \\
\left. 0,~ \Omega_\varphi +\frac{\Delta\varphi_t 2\widetilde \omega^2 \varphi -\Delta\psi_t\omega_j \omega_{10}}{\Omega_\varphi}\right] ~, ~~~~~~~~
\end{eqnarray}
where $\Omega_{\varphi}=\sqrt{\omega_{10}^2+(2\widetilde\omega\varphi )^2}$ is the level-splitting frequency and random contributions of $\bbomega_t$ give rise to the noise-induced inter-level transitions and the renormalization of frequency $\Omega_{\varphi}$. The projection operators on the upper ($+$) and lower ($-$) levels are determined as $\hat{\cal P}_{\pm}=(1\pm \hat\sigma_z)/2$ which is similar to Eq. (6) but $\hat{\cal P}_{\pm}$ is written in the basis formed by the eigenstates determined by the  Hamiltonian (7).
%f2
\begin{figure}
\begin{center}
\includegraphics[scale=0.22]{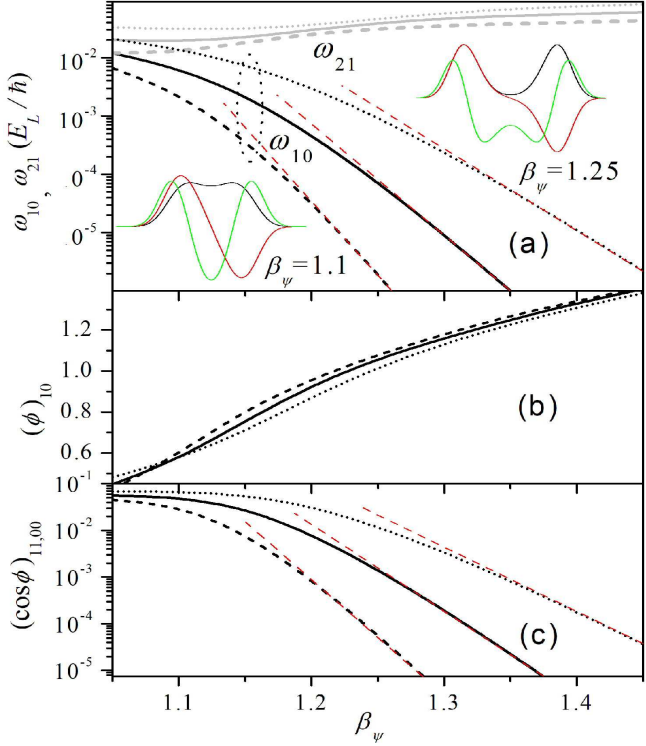}
\end{center}
\addvspace{-0.5 cm}
\caption{Parameters of Hamiltonians (5) and (7) versus control flux $\psi$. (a) Level splitting frequencies $\omega_{10}$ and $\omega_{21}$ in units $E_L /\hbar$ versus ratio $\beta_\psi =E_J\cos (\psi /2) /E_L$. Insets show wave functions $\langle\phi |n\rangle$ for $n=$0 (black), 1 (red), and 2 (green) at $4E_C /E_L =0.05$. (b) Matrix element $(\phi )_{10}$  versus $\beta_\psi$. (c) Matrix element $(\cos\phi )_{11,00}$ versus $\beta_\psi$. Red dashed lines in panels (a) and (c) show exponential asymptotics for the weak coupling regime. In all panels, the dotted, solid, and dashed curves correspond to $4E_C /E_L =0.01$, 0.005, and 0.0025, respectively. }
\end{figure}

Numerical solutions of the symmetric eigenstate problem $\hat H_{\varphi =0,\beta}\left| n\right\rangle =\varepsilon_{n}\left| n\right\rangle$ are obtained after discretization of the Hamiltonian (1) using  $N\sim 10^3$ points along $\phi$ axis and diagonalization of $N\times N$ matrix. In Fig. 2(a) we plot the dimensionless splitting frequencies $\omega_{10}$ and $\omega_{21}$ as well as the wave functions for $n=0$, 1, and 2 versus the ratio $\beta_\psi$ determined by the control flux. For the weak-tunneling  regime, if $\beta_\psi >1.2\div 1.3$ depending on $4E_C /E_L$, the frequency $\omega_{10}$ is suppressed exponentially. The dimensionless matrix elements $(\phi )_{10}$ and $(\cos\phi )_{11,00}$, which determine the characteristic frequencies $\widetilde\omega$ and $\omega_j$, are plotted in Figs. 2(b) and 2(c), respectively. Dependency of $(\phi )_{10}$ on $4E_C/E_L$ is negligible, see Fig. 2(b), and the matrix element $(\cos\phi )_{11,00}$ is exponentially suppressed at  $\beta_\psi >1.2\div 1.3$, see asymptotes in Figs. 2(a) and 2(c). Further we analyze relaxation rates versus $\omega_{10}$ and frequencies in Eq. (5). Under mapping between $\omega_j$- and $\psi$-variables the non-monotonic dependence of $\omega_j ={\rm sign}(\psi )(E_J /2\hbar )(\cos\phi )_{11,00}\sqrt{1-(\beta_\psi /\beta_0 )^2}$ on $\psi$ may be noticeable; below we use $\psi >0$ so that $\omega_j >0$.   

%%%%%%%%%%%%%%%%%%%%%%%%%%%%%%%%%%%%%%%
\section{Incoherent resonant tunneling }
We evaluate here the rate of the incoherent resonant tunneling  between $l$- and $r$-wells based on the Bloch equation with the random frequency $\bomega_t$ given by Eq. (5). This system is transformed into the integro-differential equation for the redistribution of population between wells. The $2\times 2$ density matrix takes form $\hat\rho_t =(1+{\bf S}_t\cdot{\hat\bsigma})/2$ with $k=l,r$ and the population in $k$th well $n_k ={\rm tr}\hat P_k\hat\rho_t$ is determined through the projection operators, $\hat P_k$, given by Eq. (6). At $t>0$ the Bloch vector ${\bf S}_t$ is governed by the standard equation
%8
\begin{equation}
\frac{d{\bf S}_t}{dt} =\left[{\bomega}_t\times {\bf S}_t\right] ~, ~~~~ {\bf S}_{t=0}=(\mp 1,0,0)
\end{equation}
with the normalization condition $|{\bf S}_t |=1$ and the initial vector ${\bf S}_{t=0}$, where $S_{t=0x}\equiv S_0 =-1$ or $+1$  corresponds to the state localized in the $l$- or $r$-wells, respectively. In order to reduce the system (8) into equation for $S_{tx}$ we  write $S_{ty,z}$ through the integrals $\int_0^t {dt'}\cdots S_{t'x}$ as
%9
\begin{equation}
\left|\begin{array}{*{20}c}  S_{ty} \\ S_{tz}\end{array}\right| =\int\limits_0^t{dt'} \left| \begin{array}{*{20}c} \cos\theta_{tt'}  \\ \sin\theta_{tt'}\end{array} \right|\Omega_{t'z}S_{t'x} ~. 
\end{equation}
After substitution of $S_{ty,z}$ into the X-component of Eq. (8) one obtains the exact integro-differential equation
%10
\begin{equation}
\frac{dS_{tx}}{dt}+\omega_{tz}\int\limits_0^t{dt'}\omega_{t'z}\cos\theta_{tt'}S_{t'x} =0  
\end{equation}
with the phase factor $\theta_{tt'}=\int_{t'}^t{d\tau}\omega_{\tau ,x}$. 

Below we separate the averaged part of the Bloch vector, $\left\langle{S_{tx}}\right\rangle$, in Eq. (10) and consider equation
%11
\begin{eqnarray}
\frac{d\left\langle {S_{tx} } \right\rangle}{dt}\!\! +\!\! \int\limits_0^t\!\!  {dt'}K_{tt'}
\left\langle{S_{t'x}}\right\rangle\!\!  = 0 ,  ~ K_{tt'}\!\!  =\left\langle {\omega_{tz} \omega _{t'z} \cos\theta_{tt'} } \right\rangle , ~~ 
\end{eqnarray}
where the averaged kernel $K_{tt'}$ should be written through the correlators (3). We restrict ourselves by the weak-fluctuation approximation when $\delta S_t =S_{tx}-\langle{S_{tx} }\rangle$ can be omitted, see Sect. V A. Straightforward averaging of the kernel $\langle K_{tt'}\rangle \equiv K_{\Delta t}$ gives 
%12
\begin{eqnarray}
K_{\Delta t}=\! e^{-\Gamma_{\Delta t}}\! \left(\cos\theta_{\Delta t} {\cal A}_{\Delta t} -\sin\theta_{\Delta t}{\cal B}_{\Delta t} \right) , ~~~~ \\
{\cal A}_{\Delta t}\!\! =\!\omega_{10}^2\!\! +\! \omega_j^2\!\! \left( \alpha _{tl} w_{\omega_m\Delta t}\! - \! W_{\Delta t}^2 \right) ,~ {\cal B}_{\Delta t}\!\! =\! 2\omega_j \omega_{10} W_{\Delta t}  \nonumber
\end{eqnarray}
with the noiseless phase $\theta_{\Delta t}=2\varphi\widetilde \omega \Delta t$ caused by tilt flux, see Appendix for details. This kernel is dependent on the energies $E_{L,C,J}$ through the frequencies introduced by Eq. (5), on the noise strengths $\alpha_{tl,j,c}$ and on $\omega_{m}$ from Eq. (3), as well as on the external fluxes, $\varphi$ and $\psi$. The decrement of exponential damping, $\Gamma_{\Delta t}$, and the factor $W_{\Delta t}$ in ${\cal A}_{\Delta t}$ and ${\cal B}_{\Delta t}$ of Eq. (12) are determined by the integrals
%13
\begin{subequations}
\begin{eqnarray}
\frac{\Gamma_{\Delta t}}{2\alpha_{tl}}\!\! =\!\! \widetilde\omega^2 \!\!\! \int\limits_0^{\Delta t}\!\! {d\tau}\!\!\! \int\limits_0^{\Delta t}\!\! {d\tau '}\! w_{|\omega_m\Delta\tau |}\!\! =\!\! (\widetilde \omega \Delta t)^2\!\! \left(\!\! {w_{|\omega _m \Delta t|}\!\! +\!\!\frac{3}{{2\pi }}} \!\!\right) , \\
\frac{W_{\Delta t}}{2\alpha_c}\! =\widetilde\omega\int_0^{\Delta t}\!\! {d\tau}w_{|\omega_m\tau |}\! =\! \widetilde\omega \Delta t \left(\! {w_{|\omega_m \Delta t|}\! +\!\frac{1}{\pi}}\!\right) . ~~     
\end{eqnarray}
\end{subequations}
In order to write the explicit expressions here, we used the definition of $w_{|\omega_m\Delta\tau |}$ given by Eq. (3) and performed the integrations by parts over $\tau$ and $\tau'$. Under these integrations we replaced the upper limit $\omega_M \Delta t$ by $\infty$.

According to Eq. (6), transient evolution of the population in left [or right] well is governed by $(1-\left\langle S_{tx}\right\rangle )/2$ [or $(1+\left\langle S_{tx}\right\rangle )/2$]. The Laplace transform of the averaged Bloch vector, $S_\zeta$, is obtained from the integro-differential equation (10) with the use of the convolution theorem as follows $S_\zeta =S_0 /(\zeta +K_\zeta )$, so that $\left\langle S_{tx}\right\rangle$ takes form:
%14
\begin{equation}
\left\langle S_{tx}\right\rangle =S_0\int_C \frac{d\zeta}{2\pi i}\frac{e^{\zeta t}}{\zeta +K_\zeta} \propto e^{-\nu_T t} ,~~ {\rm if} ~ \nu_T t\gg 1 .
\end{equation}
The relaxation rate at tail of tunneling decay, $\nu_T$, is obtained as a pole of the complex integral (14), where the contour $C$ is around complex half-plane ${\rm Im}\zeta <+0$. Such a pole is determined by the equation $\zeta +K_\zeta =0$ at $\zeta\to 0$. As a result, the rate $\nu_T \approx K_{\zeta\to 0}$ is given by the integral
%15
\begin{equation}
\nu_T \approx \!\int_0^\infty d\Delta t e^{-\Gamma_{\Delta t}}\! \left(\cos\theta_{\Delta t} {\cal A}_{\Delta t} -\sin\theta_{\Delta t}{\cal B}_{\Delta t} \right) , ~ 
\end{equation}
where the time scales $\widetilde \omega\Delta t\leq{\rm const}/\sqrt{\alpha_{tl}}$ are essential only. Under integration of Eq. (15) we use the logarithmic approximation, when the factors determined by Eq. (13) are replaced by the linear and quadratic dependencies, $W_{\Delta t}\approx 2\alpha_c\Lambda_T\widetilde \omega\Delta t$ and $\Gamma_{\Delta t}\approx 2\alpha_{tl}\Lambda_T (\widetilde \omega\Delta t)^2$. Based on Eq. (3),  the logarithmic factor $\Lambda _T$ is introduced here according to
%16
\begin{equation}
w_{|\omega_m \Delta t|}\approx\ln (\sqrt {2\alpha_{tl}\Lambda_T}~\widetilde \omega /\omega_m ) +c_T = \Lambda _T ~,
\end{equation}
so that $\Lambda_T \approx \ln (\sqrt {\alpha_{tl}}\widetilde\omega /\omega_m )+c_T\gg 1$ with $c_T\sim 1$. Straightforward integration of Eq. (15) is performed for ${\cal A}_{\Delta t}\approx \omega_{10}^2$ because $\alpha_{j,c}<\alpha_{tl}\ll 1$, see estimates below. The explicit formula for the tunneling rate versus tilt flux is given by the Gaussian peak modulated by the pre-factor which is linear-dependent on $\varphi$: 
%17
\begin{eqnarray}
\nu_T  \approx\widetilde \nu _T e^{-(\varphi /\varphi _T )^2}(1-\varphi /\varphi_c ), ~~~~~~~~~   \\
\widetilde \nu _T =\frac{\sqrt\pi\omega _{10}^2}{2\varphi _T \widetilde \omega} ,  ~~ \varphi_T =\sqrt {2\alpha_{tl}\Lambda_T} , ~~ \varphi_c =\frac{\alpha_{tl}\omega _{10}}{2\alpha_c\omega _j}\Lambda_T  .   \nonumber
\end{eqnarray}
Here we introduce the amplitude of peak at symmetric point, $\widetilde \nu _T$, the half-width of peak, $\varphi_T\ll 1$, and the asymmetry parameter, $\varphi_c$. An asymmetry of peak, with respect to the replacement $\varphi\to -\varphi$ at fixed direction of $\psi$, appears due to correlations between $tl$- and $j$-channels, $\varphi_c^{-1}\propto\alpha _c$. This effect is weak enough because $\varphi_T\ll\varphi_c$ and the shift of peak, determined by the requirement $(d\nu_T /d\varphi )_{\varphi =\varphi_0}=0$, is equal $\varphi_0 \approx \varphi_T^2 /2\varphi_c \ll \varphi_T$.
%f3
\begin{figure}
\begin{center}
\includegraphics[scale=0.22]{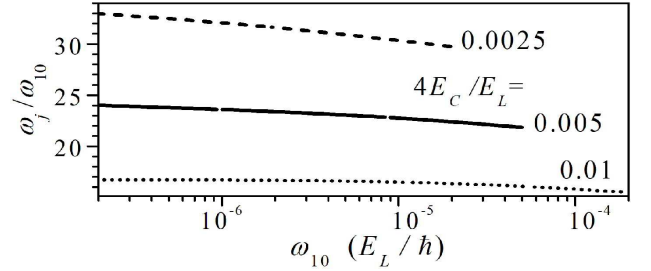}
\end{center}
\addvspace{-0.75 cm}
\caption{ The ratio $\omega_j /\omega_{10}=\varphi_0 /2\alpha_c$ determining the shift of peak $\varphi_0$,  versus gap frequency $\omega_{10}$, which  corresponds the interval $1.2\leq\beta_\psi \leq 1.45$. Different $4E_C /E_L$ are marked. }
\end{figure}
 
For the estimates below we use a typical 1/f noise level  $\leq 4\times 10^{-11}\Phi_0^2/$Hz determined by the flux-flux spectral function at 1 Hz, \cite{14,15} so that $\alpha_{tl,j,c}\leq 10^{-8}$. The half-width of peak, $\varphi_T$, is only determined by the level of noise in the $tl$-channel and by the ratio $\widetilde\omega /\omega_m$, through $\Lambda_T$. Allowing for $\Lambda_T\sim 10\div 15$, one obtains $\varphi_T\sim (4.5\div 5.5)\times 10^{-4}$, or in the  dimensional units the maximal half-width of peak is $\sim 70\div 85~\mu\Phi_0$. According to Fig. 3 the maximal shift of peak is determined by the ratio $\omega_j /\omega_{10}\sim 15\div 30$, so that $\varphi_0\approx 2\alpha_c \omega _j /\omega _{10}\leq (3\div 6)\times 10^{-7}$ or $\leq 0.1 ~\mu\Phi_0$. Since $\widetilde\omega\sim 1.2 E_L /\hbar$ is weakly ($<$ 20\% ) dependent on the barrier-control variable $\beta_\psi$ or $\omega_{10}$ [Fig. 2(b)], the amplitude $\widetilde \nu _T$ is approximately proportional to $\omega_{10}^2$, i.e. the rate increases, when overlap of the left and right wave functions increases. If $\omega_{10}/2\pi \simeq$0.5 MHz for the device with $E_L /h\approx 250$ GHz, one obtains $\widetilde \nu _T\approx 8.4\div 10.3$ ms$^{-1}$ at $\alpha_{tl}\simeq 10^{-8}$ and the maximal tunneling rate decreases with a noise level as $\widetilde \nu_T\propto\alpha_{tl}^{-1/2}$. The numerical estimates performed show that the mechanism suggested gives an essential contribution to the resonant tunneling reported in \cite{15} but a shape of peak differs from the experimental data, see discussion in Sect. VI. 

%%%%%%%%%%%%%%%%%%%%%%%%%%%%%%%%%%%%%%%
\section{Weak qubit-noise coupling }
We consider the weak qubit-noise coupling, when the gap frequency $\Omega_{\varphi}$ exceeds the noise-induced modulation of levels and it is convenient to use the tilt qubit frame described by the Hamiltonian (7). The $2\times 2$ density matrix $\hat\rho_t =(1+{\bf R}_t\cdot{\hat\bsigma})/2$ is written below through the Bloch vector ${\bf R}_t$ which is governed by the equation $d{\bf R}_t /dt =\left[{\bbomega}_t\times {\bf R}_t\right]$ with the normalization condition $|{\bf R}_{t=0}|=1$. Introducing the circular components $R_{t\pm}=(R_{tx}\pm iR_{ty} )/2$ we arrive to the system
%18
\begin{equation}
\left\{ {\begin{array}{*{20}c}
   {\left( {d/dt \mp i\Omega _{tz} } \right)R_{t \pm }  \pm i\Omega_{tx}R_{tz}/2  = 0}  \\
{dR_{tz}/dt + i\Omega_{tx}(R_{t+}\! -\! R_{t-} )= 0} \end{array}} \right.  ,
\end{equation}
where $R_{t-}=R_{t+}^*$ because $\hat\rho_t =\hat\rho_t^+$. We consider the dissipative dynamics of Z- and $\pm$-components of ${\bf R}_t$ separately. Using the initial condition $R_{t=0\pm}=0$ and expressing $R_{t\pm}$ through $R_{tz}$ one obtains the exact integro-differential equation for the Z-component of the Bloch vector
%19
\begin{equation}
\frac{dR_{tz}}{dt}+\Omega_{tx} \int_0^t{dt'}\Omega_{t'x}\cos\Theta_{tt'} R_{t'z} =0 
\end{equation}
with the phase factor $\Theta_{tt'}=\int_{t'}^t{d\tau}\Omega_{\tau ,z}$. For $t\to 0$ we use the initial condition $R_{t=0z}=1$ which corresponds to the initial population localized at the upper level because the population numbers are $n_{\pm t}={\rm tr}\hat{\cal P}_\pm \hat\rho_t$ (the conservation requirement is $n_{+,t}+n_{-,t}=1$). Similarly, using the initial condition $R_{t=0z}=0$ and eliminating the Z-component  from the upper Eq. (18), we obtain the exact integro-differential equation for the $\pm$ circular components 
%20
\begin{equation}
{\left( {\frac{d}{{dt}}\! \mp\! i\Omega _{tz} } \right)\! R_{t \pm }  \pm \frac{{\Omega _{tx} }}{2}\!\!\int\limits_0^t\! {dt'} \Omega _{t'x} (R_{t' + }\! -\! R_{t' - } )\! =\! 0} ~.
\end{equation}
As an initial condition for Eq. (20) we use the normalization condition at $t=0$ written in the form $R_{t=0+}R_{t=0-}=1/4$. Below we restrict ourselves by the weak-fluctuation regime and consider the averaged dynamics determined by Eqs. (19) and (20).

%%%%%%%%%%%%%%%%%%%%%%%%%%%%%%%%%%%%%%%
\subsection{Relaxation of population } 
We consider the interlevel redistribution of population, $\Delta n_t\equiv n_{+t}-n_{-t}={\rm tr}\hat \sigma_z \hat\rho_t =R_{tz}$, and separate $R_{tz}$ into the averaged and random parts, $R_{tz}=\langle{\Delta n}_t\rangle +\delta n_t$, which are governed by the equations
%21
\begin{eqnarray}
\frac{d}{dt}\left| {\begin{array}{*{20}c} \langle{\Delta n}_t\rangle  \\  {\delta n_t }  \end{array}} \right| + \int_0^t {dt'} N_{t - t'} \left| {\begin{array}{*{20}c}
   \langle{\Delta n}_{t'}\rangle  \\  {\delta n_{t'} }  \end{array}} \right| = \left| {\begin{array}{*{20}c} 0  \\   {-\delta f_t }  \end{array}} \right| , ~~ \\
N_{\Delta t} =\! \left\langle \Omega _{tx} \Omega _{t'x} \cos\Theta_{tt'} \right\rangle \! \approx\!\cos (\Omega_\varphi \Delta t){\cal X}_\varphi w_{\omega_m\Delta t}/\Omega _\varphi^2  .  \nonumber 
\end{eqnarray}
Here we approximate the averaged kernel $N_{\Delta t}$ neglecting random contributions to the phase $\Theta_{tt'}\to\Omega_\varphi \Delta t$, see Eq. (A5). Based on Eqs. (3) and (7) we used the correlator $\left\langle \Omega _{tx} \Omega _{t'x} \right\rangle ={\cal X}_\varphi w_{\omega_m\Delta t}/\Omega _\varphi^2$, where 
%22
\begin{equation}
{\cal X}_\varphi\! =\! \alpha _{tl} (2\omega _{10} \widetilde \omega )^2\!  +\! \alpha _j (\omega _j \widetilde \omega \varphi )^2\! +\!\alpha_c \omega _{10} \omega _j (2\widetilde \omega )^2\varphi  ~~~
\end{equation}
is dependent on the control parameters $\varphi$ and $\psi$ (through $\beta_\psi$). Fluctuations of population ${\delta n}_t$ are governed by the non-uniform equation with the source $\delta f_t =\int_0^t {dt'}\cos (\Omega_\varphi \Delta t)\delta N_{tt'} \left\langle \Delta n_{t'}\right\rangle$ determined by a random part of kernel $\delta N_{tt'}=\Omega _{tx}\Omega _{t'x}-\left\langle \Omega _{tx}\Omega _{t'x}\right\rangle$ and we consider this contribution in Sect. V B.

After the Laplace transform of the upper line of Eq. (21) with the initial condition $\langle{\Delta n}_{t=0}\rangle =1$, we obtain the transient solution
%23
\begin{equation}
\Delta n_\zeta\! =\left[\zeta  + {\cal X}_\varphi (w_{\zeta -i\Omega _\varphi} +w_{\zeta +i\Omega_\varphi})/2\Omega_\varphi^2\right]^{-1}  ,
\end{equation}
which is written through $w_\zeta =\ln [(\zeta /\omega_m )^2 +1]/2\pi\zeta$. Similarly to Sect. III, we use the logarithmic approximation  $w_{\zeta\pm i\Omega_\varphi}\approx\mp i\left(\Lambda_1 \pm i\pi /2\right)/\pi \Omega_\varphi$ with $\Lambda_1\approx\ln (\Omega_\varphi /\omega_m )+c_1\gg 1$ and $c_1\sim 1$. The $\propto\Lambda_1$ contributions cancel each other  in Eq. (23) and $\Delta n_\zeta\approx (\zeta +\nu _1 )^{-1}$ or $\langle{\Delta n}_t\rangle =\exp (-\nu_1 t)$ describes the nonoscillating  exponential decay with the population relaxation rate $\nu _1\approx {\cal X}_\varphi /2\Omega_\varphi^3$. At the symmetric point $\varphi =0$, this rate is given by $\nu _1 |_{\varphi =0}={\cal X}_{\varphi =0} /\omega_{10}\approx 2\alpha_{tl} \widetilde \omega ^2 /\omega _{10}$, when only $tl$-channel remains essential. Under non-zero  tilt $\varphi\neq 0$, the relaxation rate is suppressed for the most part due to the factor $\Omega_\varphi^{-3}$. Using a typical half-width of peak $\sim\omega_{10}/2\widetilde\omega$, we find that the $\varphi$-dependent contributions to ${\cal X}_\varphi$ are weak because $\alpha_{j,c}\leq \alpha_{tl}$ and $\omega_j /4\widetilde\omega\ll 1$ [see Figs. 2 (b) and (c)]. As a result, the rate $\nu_1$ takes form
%24
\begin{equation}
\nu _1 \approx\frac{\widetilde \nu_1}{\left[ {1 + (\varphi /\varphi_1 )^2 } \right]^{3/2}} , ~~ \widetilde \nu_1 =\frac{2\alpha_{tl}\widetilde\omega ^2}{\omega_{10}} ,  ~~
\varphi_1 =\frac{\omega_{10}}{2\widetilde \omega}
\end{equation}
and, in analogy to the incoherent tunneling regime, the relaxation of population is only governed by the noise in the $tl$-channel, $\nu_1\propto\alpha_{tl}$.
%f4
\begin{figure}
\begin{center}
\includegraphics[scale=0.22]{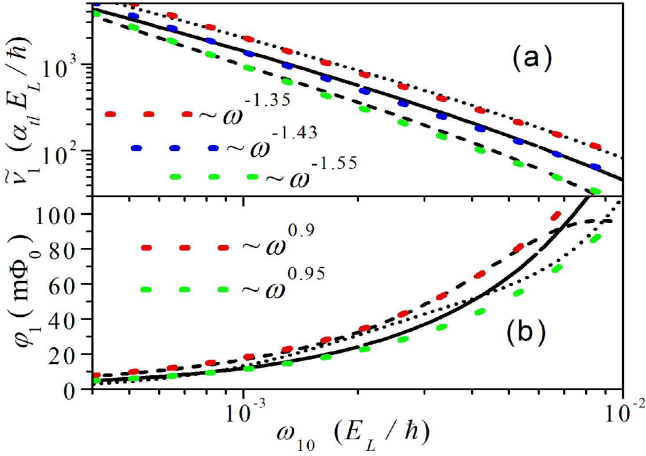}
\end{center}
\addvspace{-0.5 cm}
\caption{Parameters determining relaxation of population in Eq. (24). (a) Maximal rate at symmetric point $\widetilde\nu_1 =\nu_1 |_{\varphi =0}$ versus gap frequency $\omega_{10}$ corresponding the spectral region 0.1$\div$2.5 GHz at $E_L /h=$250 GHz. Color dots  approximate $\widetilde\nu_1$ by $\propto\omega^{-q}$ dependencies with $q=1.35\div 1.55$. (b) Half-width of the rate $\varphi_1$ for the same spectral region. Color dots show near-linear approximations. In both panels, the dotted, solid, and dashed curves correspond to $4E_C /E_L =0.01$, 0.005, and 0.0025, respectively. }
\end{figure}

Apart from the tails of the rate $\nu _1 /\widetilde\nu_1\leq 10^{-2}$ versus $\varphi$, where the $\propto\varphi ,~\varphi^2$ contributions to ${\cal X}_\varphi$ are not negligible, the shape of the peak is determined by the amplitude and the half-width, $\widetilde \nu_1$ and $\varphi_1$ (we neglect a weak asymmetry of $\nu_1$ aroused from $\propto\alpha_c\varphi$ contributions). According to Eq. 2 (b), $\widetilde \omega$ varies from $\sim$0.4 to $\sim$1 over the weak-coupling region due to  the softness of barrier, so that the dependency of $\widetilde \nu_1$ on the gap frequency $\omega_{10}$ {\it differs from} $\propto\omega_{10}^{-1}$ as it is shown in Fig. 4(a) for the different ratios $4E_C /E_L$. The weak-coupling regime takes place under the requirement $\nu_1 /\Omega _\varphi\ll 1$ which is transformed into the condition $2\alpha_{tl}(\widetilde\omega /\omega_{10})^2\ll 1$ at the symmetric point $\varphi =0$. If $\alpha_{tl}\leq 10^{-8}$ this condition is satisfied and for $E_L /h\approx$250 GHz the maximal rate is $\widetilde \nu_1 \approx (0.07\div 0.16 )~\mu$s$^{-1}$ at $\omega_{10}/2\pi =$0.5 GHz. The half-width of the peak at the same $E_L /h$ and $\omega_{10}/2\pi$ is about $\varphi_1\approx 22\div 33$  m$\Phi_0$ and  $\varphi_1$ shows near-linear dependencies on $\omega_{10}$, as it is plotted in Fig. 4(b).

%%%%%%%%%%%%%%%%%%%%%%%%%%%%%%%%%%%%%%%%%%%%%
\subsection{Decoherence rate}
The decoherence process of the off-diagonal part of density matrix is described by the circular components of the Bloch vector governed by Eq. (20). Separating rotation with the frequency $\Omega_\varphi$ according to $R_{t \pm}=\exp (\pm i\Omega_\varphi t)r_{t\pm}$ and neglecting the high-frequency contributions [$\propto\exp (\pm i\Omega_\varphi (t+t'))$], we arrive to the closed equations for the slow amplitudes $r_{t\pm}$:
%25
\begin{equation}
\left(\! {\frac{d}{dt} \mp i\Delta \Omega _{tz} }\!\! \right)r_{t \pm }\! +\!\frac{{\Omega _{tx} }}{2}\!\!\int_0^t\!\! {dt'} \Omega _{t'x} e^{ \mp i\Omega _\varphi  (t - t')} r_{t' \pm }\!\!  =\!\! 0 ~.
\end{equation}
Because $r_{t-}=r_{t+}^*$, we consider only the $+$ component $r_{t+}\equiv r_t$ . Within the weak-fluctuation regime, we separate the equations for the averaged and random parts of the amplitude, $r_t=\left\langle r_t\right\rangle +\delta r_t$, as follows
%26
\begin{eqnarray}
\frac{d}{dt}\left| {\begin{array}{*{20}c}
   {\left\langle {r_t} \right\rangle }  \\
   {\delta r_t }  \\
\end{array}} \right| -i\left| {\begin{array}{*{20}c}
   {\left\langle {\Delta \Omega _{tz} \delta r_t } \right\rangle }  \\
   {\Delta \Omega _{tz} \left\langle {r_t} \right\rangle }  \\
\end{array}} \right|   ~~~~~~~   \\
+ \frac{1}{2}\int_0^t {dt'} e^{-i\Omega _\varphi  (t - t')} \left| {\begin{array}{*{20}c}
   {\left\langle {\Omega_{tx} \Omega_{t'x} } \right\rangle } \\
   {\delta N_{tt'} }  \\
\end{array}} \right|\left\langle r_{t'} \right\rangle  = 0 ~.  \nonumber
\end{eqnarray}
Here the random part of kernel $\delta N_{tt'}$ gives vanishing contribution to $\left\langle {\Delta \Omega _{tz} \delta r_{t \pm } } \right\rangle$, because it contains an average of one or three random factors. As a result, $\left\langle {r_{t \pm } } \right\rangle $ is governed by the equation
%27
\begin{eqnarray}
\frac{{d\left\langle {r_t} \right\rangle }}{{dt}} + \int_0^t {dt'} \left\langle {\Delta \Omega _{tz} \Delta \Omega _{t'z} } \right\rangle \left\langle {r_{t'}} \right\rangle ~~~~ \\
+ \frac{1}{2}\int_0^t {dt'} e^{-i\Omega _\varphi  (t - t')} \left\langle\Omega _{tx} \Omega _{t'x} \right\rangle \left\langle {r_{t'}}\right\rangle\simeq 0  \nonumber
\end{eqnarray}
with the initial condition $\left\langle r_{t=0}\right\rangle =r_0$ where $r_0$ is normalized by the requirement $r_0r_0^* =1/4$. The first and second integral terms here describe the long- and short-scale contributions to the temporal evolution.

After the Laplace transform of Eq. (27), we obtain the transient solution
%28
\begin{equation}
r_{\zeta}=r_0 \left[ {\zeta +{\cal Z}_\varphi w_\zeta /\Omega _\varphi^2 +{\cal X}_\varphi w_{\zeta  +i\Omega _\varphi  } /2\Omega _\varphi^2} \right]^{-1} .
\end{equation}
Here the correlator $\left\langle\Delta\Omega _{tz} \Delta \Omega_{t'z}\right\rangle ={\cal Z}_\varphi w_{\omega_m\Delta t}/\Omega _\varphi^2$ is written through
%29
\begin{equation}
{\cal Z}_\varphi\! =\! \alpha _{tl} (2\widetilde\omega^2 \varphi )^2\!  +\! \alpha _j (\omega_j\omega_{10})^2\! -\!\alpha_c\omega_{10} \omega _j (2\widetilde \omega )^2\varphi ~,
\end{equation}
which is similarly $\left\langle\Omega _{tx}\Omega_{t'x}\right\rangle$ in Eq. (22). We consider below a slow-varied part of $\left\langle r_t\right\rangle$, when $|\zeta |\sim\nu_2 \ll\Omega_\varphi$ with the decoherence rate $\nu_2$, and apply the logarithmic approximation, when $w_\zeta$ is replaced by $[\Lambda_2 +i{\rm arg}(\zeta )]/\pi\zeta$ with $\Lambda_2\approx\ln (\nu_2 /\omega_m )+c_2$ with $c_2\sim 1$. In analogy to Sect. IV A, $w_{\zeta +i\Omega_\varphi}$  should be replaced by $(1/2 -i\Lambda_1 /\pi )/\Omega_\varphi$ with $\Lambda_1 >\Lambda_2$ because of $\Omega_\varphi\gg\nu_2$. Using these replacements, one can search the poles of the solution $r_{\zeta}$ from the quadratic equation:
%30
\begin{equation}
\zeta ^2  + \zeta \nu_1 \left( 1/2-i\Lambda_1 /\pi\right) +\nu_\varphi^2 [ \Lambda_2  + i\arg (\zeta )]/\pi \approx 0 ~,
\end{equation}
where we introduce the characteristic rate $\nu_\varphi\equiv\sqrt{{\cal Z}_\varphi}/\Omega_\varphi$ and use $\nu _1\approx {\cal X}_\varphi/2\Omega_\varphi^3$ from Sect. IV A. This equation is transformed into $(\zeta -\zeta_+ )(\zeta -\zeta_- )\approx 0$ with a two poles $\zeta_\pm$ determined by the relation:
%31
\begin{eqnarray}
\zeta_\pm   =  - \nu _1 (1/2 - i\Lambda _1 /\pi )/2 ~~~~~~~~~~~~  \\
 \pm \sqrt {\nu _1^2 (1/2 - i\Lambda _1 /\pi )^2 /4 - \nu _\varphi ^2 [\Lambda _2  + i\arg (\zeta )] /\pi }  \nonumber
\end{eqnarray}
and it is not an explicit solution because of the factor $\arg{(\zeta )}$. Within the logarithmic accuracy, $\Lambda_{1,2}\gg\pi /2$, the poles are near the imaginary axis and $\arg (\zeta_\pm )\approx\pm\pi /2$. 

Thus the poles (31)  describe {\it a two mode behavior} of the decoherence process. This is because of a different character of the integral contributions in Eq. (27): while the XX-term is cutting off at $t\sim\Omega_\varphi^{-1}$, the ZZ-term shows a long-time memory. We obtain the poles as $\zeta_\pm \equiv i\Delta\Omega_\pm -\nu_\pm$  with the renormalizations of the gap frequency $\Delta \Omega_\pm$ and the decoherence rates $\nu_\pm$ given by: 
%32
\begin{eqnarray}
\Delta \Omega _ \pm   \approx \nu _1 \Lambda _1 /2\pi  \pm \sqrt {(\nu _1 \Lambda _1 /2\pi )^2  + \nu _\varphi ^2 \Lambda _2 /\pi } ~, \nonumber  \\
\nu_ \pm \approx \frac{{\nu _\varphi ^2  + \nu _1 |\Delta \Omega _ \pm  |}}{{4\sqrt {(\nu _1 \Lambda _1 /2\pi )^2  + \nu _\varphi ^2 \Lambda _2 /\pi } }}  ~. ~~~~~~~~
\end{eqnarray}
After the inverse Laplace transform one obtains
%33
\begin{equation}
\frac{\langle {r_t }\rangle}{r_0}\!\! \approx\!\! \frac{{(i\Delta \Omega_+\!\! -\! \nu _ +  )e^{(i\Delta \Omega_+\! -\! \nu _ +\! )t}\!\!  -\! (i\Delta \Omega_-\!\! -\! \nu _- )e^{(i\Delta \Omega_-\!\! -\!\nu_- \! )t} }}{i(\Delta \Omega_+ -\Delta \Omega_- ) - (\nu_+ -\nu_- )} ,
\end{equation}
i.e. an oscillating temporal evolution of $\langle {r_t }\rangle$ is described through $\Delta \Omega_\pm$ and $\nu_\pm$ determined by Eq. (32).
%f5
\begin{figure}
\begin{center}
\includegraphics[scale=0.21]{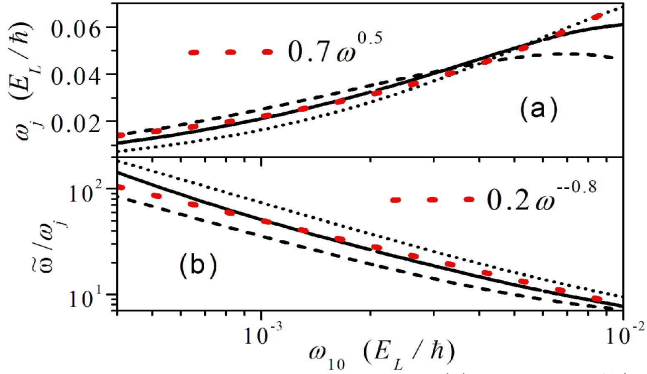}
\end{center}
\addvspace{-0.8 cm}
\caption{ Characteristic frequency $\omega_j$ (a) and ratio  $\widetilde \omega /\omega_j$ (b), which determine the decoherence process, versus gap frequency $\omega_{10}$ for the same conditions as in Fig. 4 (a). Asymptotics are shown by red dots. }
\end{figure}  
Shape of the damping oscillations in (33) is determined by the rates $\nu_1$ given by Eq. (24) and the characteristic rate $\nu _\varphi$ given by
%34
\begin{equation}
\frac{\nu _\varphi }{\sqrt{\alpha_j}~\omega_j}= \sqrt {\frac{{1 + \frac{{\alpha _{tl} }}{{\alpha _j }}\left( {\frac{{\widetilde\omega \varphi }}{{\omega _j \varphi _1 }}} \right)^2  - \frac{{2\alpha _c }}{{\alpha _j }}\frac{\widetilde\omega \varphi}{{\omega _j \varphi _1 }}}}{1 + (\varphi /\varphi _1 )^2 }} ~.
\end{equation}
%f6
\begin{figure}
\begin{center}
\includegraphics[scale=0.22]{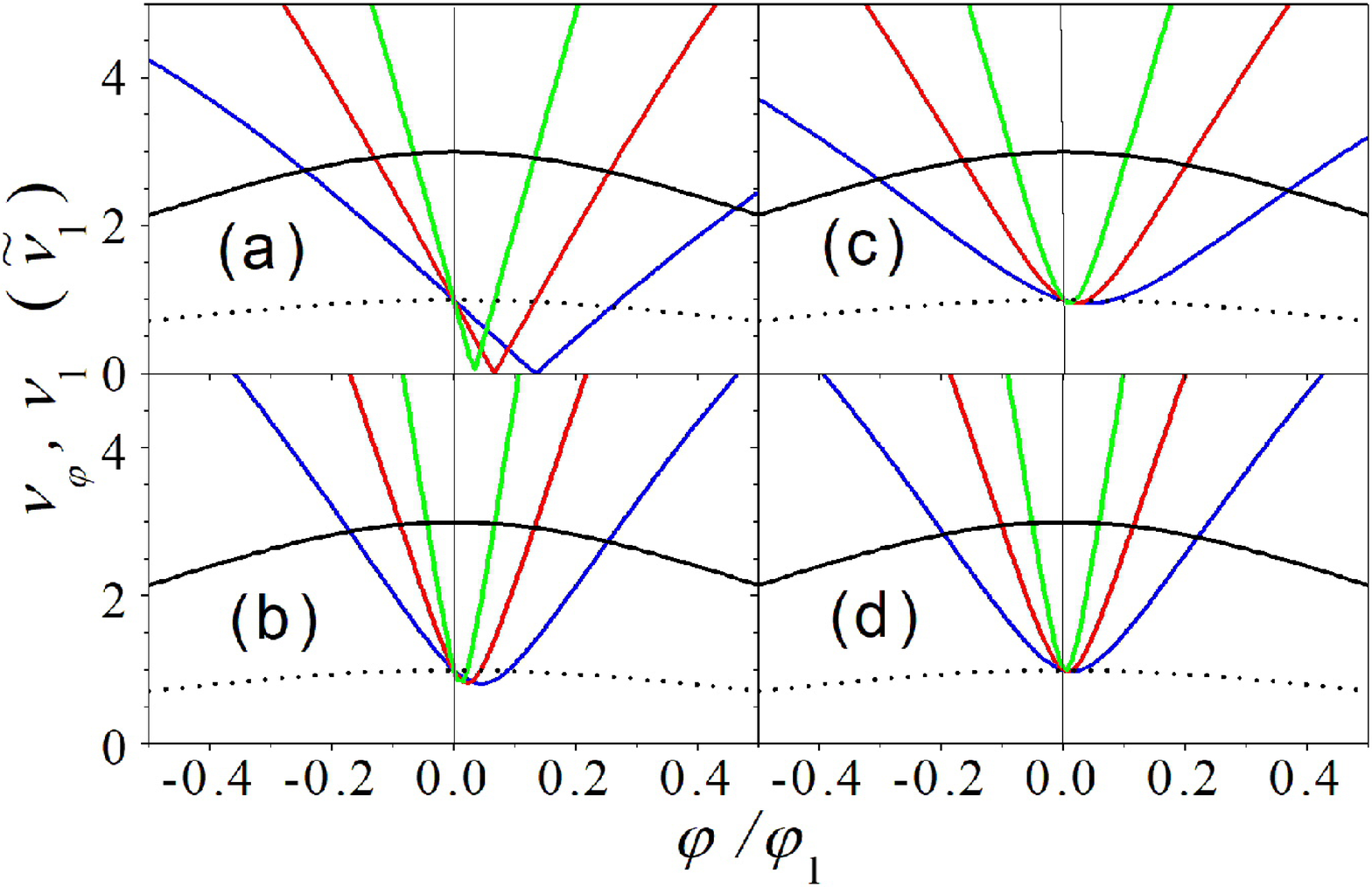}
\end{center}
\addvspace{-0.8 cm}
\caption{ Normalized rates $\nu_\varphi /\widetilde\nu_1$ (color dips) and $\nu_1 /\widetilde\nu_1$ (black) versus tilt $\varphi /\varphi_1$ plotted for $\omega_j >0$ (the case $\omega_j <0$ corresponds $\varphi\to -\varphi$). $\nu_\varphi$ is plotted for $\widetilde \omega /\omega_j =$7.5 (blue), 15 (red), and 30 (green) with different relative contributions of $tl$- and $c$-channels in panels (a)-(d): $[\alpha_{tl}/\alpha_j ,\alpha_c /\alpha_j ]=$[1,1] (a), [3,1] (b), [1,0.1] (c), and [3,0.1] (d). Black curves show $\nu_1 /\widetilde\nu_1$ for $\widetilde\nu_1 /\nu_{\varphi =0}=$1 (dashed) and 3 (solid). }
\end{figure}

This rate at symmetric point, $\nu_{\varphi =0}=\sqrt{\alpha_j}\omega_j$, and the tilt dependency of $\nu_\varphi$ are written through $\omega_j$ and $\widetilde\omega /\omega_j$ shown versus $\omega_{10}$ in Figs. 5(a) and 5(b), respectively. In contrast to $\nu_1\propto\alpha_{tl}$, we obtain $\nu_\varphi \propto \sqrt{\alpha_j}$ and the both contributions may be essential even if $\alpha_j\ll\alpha_{tl}$. There are two limiting cases, with ZZ- or XX-correlators dominate in Eq. (27), when the renormalization frequencies and the decoherence rates take forms:
%35
\begin{subequations}
\begin{eqnarray}
\Delta \Omega _ \pm\!\approx\!\pm \nu _\varphi  \sqrt {\frac{{\Lambda _2 }}{\pi }} , ~~~~\nu _ \pm\!\approx\! \frac{\nu _\varphi}{4}\sqrt {\frac{\pi}{\Lambda_2}} , ~{\rm if} ~ \nu _\varphi\!\gg\! \nu _1 , ~~    \\
\Delta\Omega_+\!\!\approx\! \nu_1\frac{\Lambda_1}{\pi} , ~ \nu_+\!\!\approx\!\frac{\nu_1}{2}, ~ \Delta\Omega_-\!\!\approx\!\nu_-\!\!\approx\! 0, ~ {\rm if} ~ \nu _\varphi\!\ll\!\nu _1 . ~~
\end{eqnarray}
\end{subequations}
Here the imaginary and real parts of poles are connected as $|\Delta\Omega_\pm |= (4\Lambda_2 /\pi )\nu_\pm$ if $\nu_\varphi\!\gg\! \nu _1$, or as $\Delta\Omega_+ = (2\Lambda_1 /\pi )\nu_+$ if $\nu_\varphi\!\ll\! \nu _1$, i.e. they differ by the scale factors $\propto\Lambda_{1,2}/\pi\gg 1$. Note, that $\nu_\varphi$ saturates at $\nu_{max}=\sqrt{\alpha_{tl}}\widetilde\omega$ and increases $\propto\varphi$ before saturation, if $1>|\varphi |/\varphi_1 |>\omega_j /\widetilde\omega$. An interplay between these regimes is significant in the region where $\widetilde\nu_1 /\nu_{\varphi =0}=\alpha_{tr}\widetilde\omega /\sqrt{\alpha_j}\omega_j\varphi_1\sim 1$. Fig. 6 compare shapes of the sharp dips of $\nu_\varphi$ and the slow peaks of $\nu_1$ for different levels of noise and different ratios $\widetilde\nu_1 /\nu_{\varphi =0}$. For the weakly-correlated noises with $\alpha_c\ll\alpha_{tl,j}$, see Figs. 6 (c) and (d), there are near-symmetric dips but for $\alpha_c\sim\alpha_j$ the dips around $\varphi\sim 0$ are asymmetric enough and are shifted to the right if $\omega_j >0$, see Figs. 6(a) and(b). Moreover, for the fully-correlated identical noises with $\alpha_{tl}=\alpha_j =\alpha_c$, one obtains $\nu_\varphi =0$ at $\varphi =(\widetilde\omega /\omega_j )\varphi_1$, see Fig. 6(a). This peculiarity is caused by {\it the destructive interference} of noises in $tl$- and $j$-channels and it becomes essential if $\alpha_{j,c}/\alpha_{tl}> 0.5$, see Fig. 8 below.
%f7
\begin{figure}
\begin{center}
\includegraphics[scale=0.21]{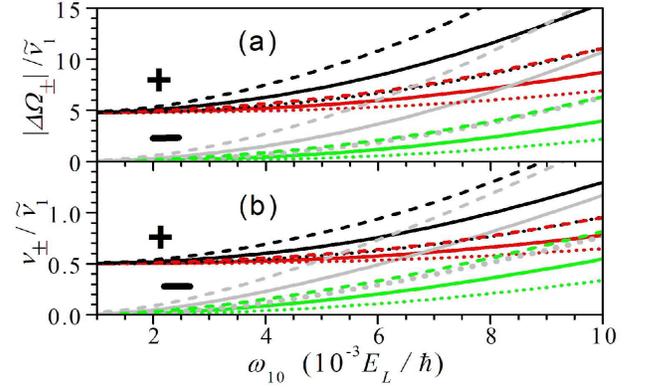}
\end{center}
\addvspace{-0.8 cm}
\caption{ Normalized frequencies $|\Delta\Omega_\pm |/\widetilde\nu_1$ (a) and decoherence rates $\nu_\pm /\widetilde\nu_1$  (b) at the symmetric point $\varphi =0$ versus gap frequency $\omega_{10}$ for the same $4E_C /E_L$ as in Figs. 4 and 5. Coupling strengths are $\alpha_{tl}=10^{-8}$ and $\alpha_j /\alpha_{tl}=$0.25 (black and gray curves) or 0.05 (red and green curves). $\pm$ modes are marked and logarithms are chosen as $\Lambda_1\simeq 15$ and $\Lambda_2\simeq 10$. } 
\end{figure}

At $\varphi =0$ the interplay of the long- and short-scale ($\propto\nu_{\varphi =0}$ and $\propto\nu_1$)  contributions in Eq. (32)  gives rise to the dependencies of $\Delta\Omega_\pm$ and $\nu_\pm$ which increase with $\omega_{10}$ as it is shown in Fig. 7. Similarly to Figs. 4 and 5, the growth of $\Delta\Omega_\pm$ and $\nu_\pm$ is dependent on $4E_C /E_L$ and on the noise levels, $\alpha_{tl}$ and $\alpha_j$. Finally, in Fig. 8 we plot $\Delta\Omega_\pm$ and $\nu_\pm$ versus tilt flux, $\varphi /\varphi_1$, for different gap frequencies at $4E_C /E_L =0.005$; there is the same behavior of dips with numerical variations up to 2 times if $4E_C /E_L$ varies over 0.01$\div$0.0025. Thus, the ratio $\Delta\Omega_\pm /\nu_\pm$ is about ten and contribution of the "$-$" mode in Eq. (33) is negligible around $\varphi =0$. If $\varphi /\varphi_1 >0.2$, the decoherence rate and  $\Delta\Omega_\pm$ increase fast (up to ten times at $\varphi /\varphi_1\sim 1$, if the gap frequency $\leq$1 GHz for the parameters used above).
%f8
\begin{figure}
\begin{center}
\includegraphics[scale=0.21]{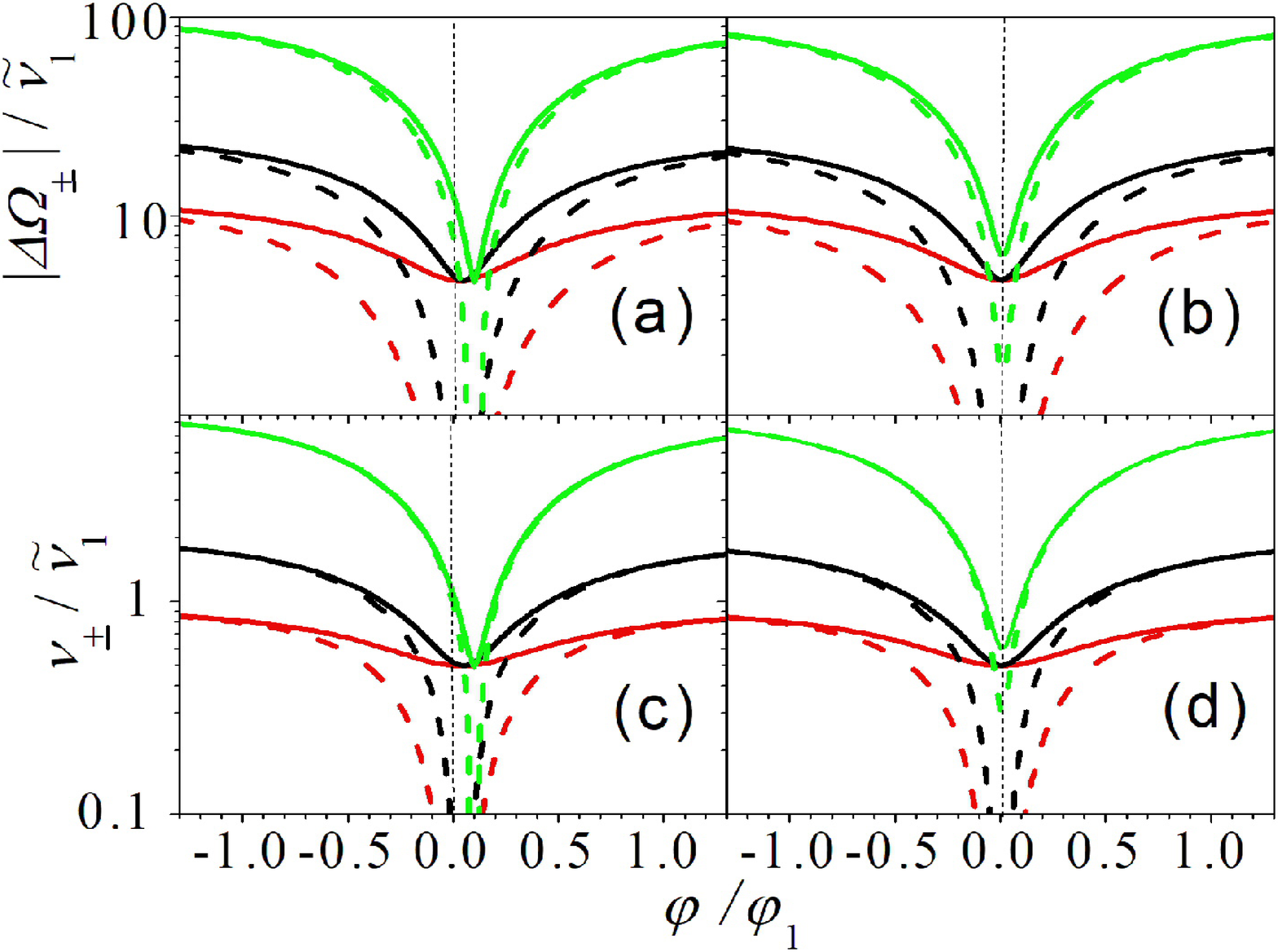}
\end{center}
\addvspace{-0.8 cm}
\caption{ Normalized frequencies $|\Delta\Omega_\pm |/\widetilde\nu_1$ (a, b) and decoherence rates $\nu_\pm /\widetilde\nu_1$  (c, d) at $4E_C /E_L =0.005$ versus tilt $\varphi /\varphi_1$ for the gap frequencies $\omega_{10}\approx 1.2\times 10^{-3}$ (green), $2.3\times 10^{-3}$ (black), and $6.5\times 10^{-3}$ (or 0.3, 0.57, and 1.6 GHz at $E_L /h=$250 GHz). Solid and dashed curves show $+$ and $-$ modes, respectively. Left (a, c) and right (b, d) panels correspond $[\alpha_j /\alpha_{tl},\alpha_c /\alpha_{tl}]=$[1,1] and [0.1,0.1], respectively. }
\end{figure}

%%%%%%%%%%%%%%%%%%%%%%%%%%%%%%%%%%%%%%%
\section{Effect of fluctuations  } 
Finally, we examine the random addendums to the averaged Bloch vectors considered in Sects. III and IV. We analyze the two-point correlations of fluctuations and determine the conditions for applicability of the averaged dynamics in the cases under consideration. These conditions are collected in the last paragraph of each subsection.

%%%%%%%%%%%%%%%%%%%%%%%%%%%%%%%%%%%%%%%
\subsection{Random contributions to incoherent tunneling }
Here we consider random part of the Bloch vector $\delta S_t =S_{tz}-\left\langle{S_t }\right\rangle$ governed by the linearized equation
%36
\begin{equation}
\frac{d\delta S_t}{dt}+\int\limits_0^t {dt'}K_{\Delta t} \delta S_{t'}  =-\int\limits_0^t {dt'} \delta K_{tt'} \left\langle {S_{t'x} }\right\rangle  \equiv\delta F_t ~,
\end{equation}
which is obtained after subtraction of the averaged Eq. (11) from the exact Eq. (10). Here a random source $\delta F_t$ is introduced trough the kernel $\delta K_{tt'}=\omega_{tz}\omega_{t'z}\cos\theta_{tt'} - K_{t - t'}$ and a second-order correction ($\propto\delta K_{tt'}\delta S_{t'}$) is omitted. Similar to Sect. III, Laplace transform of Eq. (36) gives the solution $\delta S_\zeta =\delta F_\zeta /(\zeta +K_\zeta )$ and the correlation function $\left\langle {\delta S_{t_1}\delta S_{t_2} }\right\rangle$ takes form:  
%37
\begin{eqnarray}
\left\langle {\delta S_{t}\delta S_{t'} }\right\rangle  = \int_C{\frac{d\zeta}{2\pi i}} \int_C {\frac{{d\zeta '}}{2\pi i}} \frac{e^{\zeta t +\zeta 't'}\left\langle \delta F_{\zeta}\delta F_{\zeta '}\right\rangle}{\left( {\zeta +K_{\zeta} } \right)\left( {\zeta '+ K_{\zeta '} } \right)}  \\
\approx\int_0^t {dt_1 } \int_0^{t'} {dt'_1 } e^{ - \nu _T (t + t' - t_1  - t'_1 )} \left\langle {\delta F_{t_1 } \delta F_{t'_1 } } \right\rangle . \nonumber
\end{eqnarray}
Here the correlator of random sources is written through $\left\langle{S_{t,x} }\right \rangle$:
%38
\begin{eqnarray}
\left\langle {\delta F_{t_1 }\delta F_{t'_1 } } \right\rangle\! =\!\! \int\limits_0^{t_1 }\! {dt_2 } \!\!\int\limits_0^{t'_1}\! {dt'_2 }\! \left\langle {\delta K_{t_1 t_2 } \delta K_{t'_1 t'_2 } } \right\rangle\! \left\langle {S_{t_2 x} } \right\rangle \!\langle {S_{t'_2 x} } \rangle ~, ~~~ \\
\left\langle\delta K_{t_1 t_2 } \delta K_{t'_1 t'_2 } \right\rangle\!\!\approx\! \omega_{10}^4\! \left\langle {\cos \theta_{t_1 t_2}\! \cos\theta_{t'_1t'_2 } } \right\rangle \!\!
-\!\! K_{_{t_1 -t_2} }\!\! K_{_{t'_1  - t'_2 } } ,  \nonumber 
\end{eqnarray}
where $\omega_{t_1 z}\omega_{t_2 z}\omega_{t'_1 z}\omega_{t'_2 z}$ is replaced by $\omega_{10}^4$ omitting the contributions $\propto\alpha_{\ldots}$.  

For the weak fluctuation regime, we approximate $\left\langle\cos\theta_{t_1 t_2} \cos\theta_{t'_1 t'_2} \right\rangle -\left\langle \cos\theta_{t_1 t_2}\right\rangle \left\langle\cos\theta_{t'_1 t'_2}\right\rangle$ by the first order contribution. Using (A.6) we obtain the four-point correlator in (38) as:
%39
\begin{eqnarray}
\left\langle {\delta K_{t_1 t_2 } \delta K_{t'_1 t'_2 } } \right\rangle  \approx \alpha_{tl} (2\widetilde \omega )^2 \omega _{10}^2 \cos \theta _{\Delta t_{12} } \cos \theta _{\Delta t'_{12} }  \nonumber  \\
\times \int_{t_2 }^{t_1}d\tau\int_{t'_2}^{t'_1}{d\tau}'w_{\omega_m |\tau -\tau '|}  ~~~~~~~
\end{eqnarray}
with $w_{\omega_m |\Delta \tau|}$ given by Eq. (3). At $\nu_T t_{1,2}\gg 1$ one can replace $\left\langle{S_{t,z} }\right\rangle$ by the exponent $\exp (-\nu_T t)$ and the correlator (37) is transformed into
%40
\begin{eqnarray}
\left\langle {\delta S_t^2 }\right\rangle\! \approx\! e^{-2\nu _T t} \alpha _{tl} (2\widetilde \omega )^2 \omega _{10}^4\!\! \int\limits_{\omega _m }^{\omega _M }\!\!\!\! {\frac{d\omega }{\pi\omega ^3}}\!\!\int\limits_0^t\!\! {dt_1}\!\! \int\limits_0^t\!\!{dt_1'}\!\!\int\limits_0^{t_1}\!\!{dt_2}\!\! \int\limits_0^{t_1'}\!\!{dt_2'} ~~~~  \nonumber \\
\times e^{-\nu _T (t_2 +t_2' -t_1 -t_1')} \cos \theta _{\Delta t_{12} } \cos \theta _{\Delta t'_{12} }[ \cos \omega (t_1 -t_t ') ~~~~   \\
 +\! \cos \omega (t_2 -t_2')\! -\! \cos \omega (t_2 -t_1') \!-\! \cos \omega (t_1 -t_2') ] ~. ~~~  \nonumber
\end{eqnarray}
Here we include only the contribution which is logarithmically divergent at $\omega_m\to 0$. After the straightforward integrations by parts in Eq. (40) we estimate the rare fluctuations contributions to tunneling as
%41
\begin{eqnarray}
\left\langle {\delta S_t^2 }\right\rangle\!  \approx \frac{{(\omega _{10} /\nu _T )^8 }}{{2(1 + \eta_\varphi ^2 )^4 }}F_{\nu _T t}^2 = \left( {\frac{{\omega _{10} }}{{\nu _T }}} \right)^8 \frac{{(\nu _T t)^2 }}{2}  ~~~~   \nonumber   \\
\times\left\{ {\begin{array}{*{20}c}  {(1 + e^{ - \nu _T t} )^2 ,} & {\eta _\varphi\ll 1}  \\ {[\cos (\eta _\varphi  \nu _T t) + e^{ - \nu _T t} ]^2 }, ~ & {\eta _\varphi\gg 1}  \\
\end{array}} \right. , ~~~~~~~~   \\
F_x\equiv x \left\{ {(1-\eta_\varphi^2 )\left[ {e^{ - x}  + \cos (\eta_\varphi  x)} \right] + 2\eta_\varphi  \sin (\eta_\varphi  x)} \right\} ,  \nonumber
\end{eqnarray}
where the tilt effect is described by the ratio $\eta_\varphi =2\widetilde\omega \varphi /\nu _T$.

Because of $\sqrt{\left\langle \delta S_t^2 \right\rangle}\propto\Lambda_T^3$ the rare fluctuation contributions lead to the linear growth $\sqrt{\left\langle \delta S_t^2 \right\rangle}\propto\nu_T t$ and restrict the averaged description of tunneling by the condition $\sqrt{\left\langle \delta S_t^2 \right\rangle}<e^{-\nu_T t}$. If $\omega _{10}/\nu_T <1$ and $\eta_\varphi\to 0$, this requirement is satisfied during the time interval $\nu_T t<\lambda\ln (\nu_T /\omega _{10})$ with $\lambda\sim 1$, i.e. the average description fails down starting at $\nu_T t\sim 3\div 5$, for typical parameters. A further evolution, when a fluctuations level should be saturating, is not described in the framework of the approach used. At $\varphi \sim \varphi_T$ when  $\eta_\varphi\gg 1$, the rare fluctuations are quenched as $1/\eta_\varphi^2$ so that the average description is valid outside of the peak.  

%%%%%%%%%%%%%%%%%%%%%%%%%%%%%%%%%%%%%%%
\subsection{Fluctuations of population redistribution }  
Similarly to the averaged solution for population in Sect. IV A, the solution for it's fluctuation part $\delta n_t$ with the zero initial condition is given by $\delta n_t\approx -\int_0^t {dt_1} e^{-\nu _1 (t-t_1 )}\delta f_{t_1}$. The correlation function $\left\langle{\delta n_t\delta n_{t'}}\right\rangle$ is written through the correlator of sources $\delta f_t$ as follows   
%42
\begin{equation}
\left\langle \delta n_t \delta n_{t'}\! \right\rangle\! \approx\!\! \int\limits_0^t\!\! {dt_1} e^{-\nu _1 (t-t_1 )}\!\!\!\int\limits_0^{t'}\!\!{dt_1'}e^{-\nu _1 (t'-t_1 ')}\!\! \left\langle \delta f_{t_1}\delta f_{t_1'}\! \right\rangle  .
\end{equation}
Here the correlator of sources is given by 
%43
\begin{eqnarray}
\left\langle \delta f_t \delta f_{t'}\! \right\rangle\!\!  \approx \!\!
\int_0^{t_1}\!\! {dt_2}\!\! \int_0^{t_1}\!\! {dt_2 '} \cos (\Omega_\varphi\Delta t_{12}) \cos (\Omega _\varphi\Delta t_{12}') \nonumber ~~~~~~ \\
\times \left\langle\delta N_{t_1 t_2}\delta N_{t'_1 t'_2}\right\rangle\left\langle\!\Delta n_{t_2 }\! \right\rangle \! \left\langle\! \Delta n_{t_2 '}\! \right\rangle ~~~~~~~~~~~~ 
\end{eqnarray}
and $\left\langle\delta N_{t_1 t_2}\delta N_{t'_1 t'_2}\right\rangle\equiv \left\langle\Omega _{t_1x} \Omega _{t_2x} \Omega _{t_1'x} \Omega _{t_2'x} \right\rangle - {\cal X}_\varphi ^2 w_{\omega _m \Delta t_{12}}$ $w_{\omega _m \Delta t_{12} '}/\Omega_\varphi^4$ describes the averaged fluctuations of kernel. The four-point correlator $\left\langle\Omega_{\ldots}\ldots\right\rangle$ is determined by Eq. (A7).

Within the logarithmic approximation, at $\varphi =0$ one obtains $\left\langle\delta N_{\ldots}\delta N_{\ldots}\right\rangle\approx (2\widetilde\omega )^4(\alpha _{tr}\Lambda_1 /\pi )^2$ and the mean-square-fluctuation of population takes form:
%44
\begin{eqnarray}
\left\langle {\delta n_t^2 } \right\rangle \approx (2\widetilde \omega )^2 \left( \alpha _{tr} \Lambda _1 /\pi \right)^2 e^{ - 2\widetilde\nu _1 t} ~~~~  \nonumber \\
\times\left\{ {{\mathop{\rm Re}\nolimits} \left[ {\int_0^t {dt_1 } \int_0^{t_1 } {dt_2 } e^{(\widetilde\nu _1  + i\omega _{10} )\Delta t_{12} } } \right]^2 } \right.     \\
 + \left. {\left| {\int_0^t {dt_1 } \int_0^{t_1 } {dt_2 } e^{(\widetilde\nu _1  + i\omega _{10} )t_1  - (\widetilde \nu _1  + i\omega _{10} )t_2 } } \right|^2 } \right\} . \nonumber
\end{eqnarray}
After the straightforward integrations over $t_{1,2}$ and averaging over period $2\pi /\omega_{10}$ ($\overline{\cos^2 (\omega_{10}t)}\to 1/2$), Eq. (44) gives
%45
\begin{equation}
\overline{\left\langle {\delta n_t^2 } \right\rangle}  \approx 2k^2\left[ {1/2 + e^{ - 2\widetilde\nu_1 t} (1 - \widetilde\nu_1 t)^2 } \right] ~,
\end{equation}
where $k=(2\widetilde\omega /\omega _{10})^2\alpha_{tr}\Lambda_1 /\pi\ll 1$ and $\overline{\left\langle {\delta n_t^2 } \right\rangle}\propto\Lambda_1^2$, i.e. the rare fluctuations contribution limits the averaged description of transient evolution. 

In analogy to Sect. V A, the averaged description is valid under the condition $\sqrt{\overline{\left\langle \delta n_t^2 \right\rangle}}<e^{-\nu_1 t}$ and at $\varphi =0$ this requirement is satisfied during the time interval $\widetilde\nu_1 t<\ln (1/k)\gg 1$. If $\varphi =\varphi_1$, we estimate $\Omega_{\varphi_1}\sim\sqrt{2}\omega_{10}$ and $\Omega_{tx}|_{\varphi \sim\varphi_1}\approx - \Delta \varphi_t\sqrt{2}\widetilde\omega$, so that $\sqrt{\overline{\left\langle \delta n_t^2 \right\rangle}}$ is different from Eqs. (44) and (45) by the numerical factor $\sim$1. As a result, the above condition for the averaged description remains valid over the peak described by Eq. (24). In addition, $\overline{\left\langle {\delta n_t^2 } \right\rangle}|_{\nu_1 t\gg 1}\approx k^2$ gives the steady-state level of fluctuations after redistribution of population. 

%%%%%%%%%%%%%%%%%%%%%%%%%%%%%%%%%%%%%%%
\subsection{Random contributions to decoherence } 
After integration of the lower Eq. (26) with the initial condition $\delta r_{t=0}=0$, the random contribution to the amplitude $r_t$ takes form   
%46
\begin{eqnarray}
\delta r_t\!\approx\! i\!\!\int_0^t\!\! {dt_1 } \Delta \Omega _{t_1 z}\!\! \left\langle {r_{t_1 } } \right\rangle  ~~~~~~ \\
- \frac{1}{2}\!\int_0^t {dt_1 }\! \int_0^{t_1 }\! {dt_2 } e^{ - i\Omega _\varphi  (t_1  - t_2 )} \delta N_{t_1 t_2 } \left\langle {r_{t_2 } } \right\rangle . \nonumber
\end{eqnarray}
Because the averages of $\propto\Delta \Omega _{\ldots}$ and $\propto\delta N_{\ldots}$ contributions are separated, the two-point correlation function is transformed into
%47
\begin{eqnarray}
\left\langle {\delta r_t \delta r_{t'}^* } \right\rangle  \approx \int_0^t {dt_1 } \int_0^{t'} {dt'_1 } \left\langle {\Delta \Omega _{t_1 z} \Delta \Omega _{t_1 'z} } \right\rangle \left\langle {r_{t_1 } } \right\rangle \left\langle r_{t_1 '}\right\rangle^*  
\nonumber \\
 + \frac{1}{4}\!\int_0^t\! {dt_1 }\! \int_0^{t_1 }\! {dt_2 }\! \int_0^t\! {dt_1 '}\! \int_0^{t_1 '}\! {dt'_2 } e^{ - i\Omega_\varphi  (t_1 -t_2 +t'_1 -t'_2 )} ~~~~~~~~  \\
\times \left\langle\delta N_{t_1 t_2}\delta N_{t'_1 t'_2}\right\rangle\! \left\langle {r_{t_2 } } \right\rangle\! \left\langle {r_{t_2 '}^* } \right\rangle ,  ~~~~~~~~ \nonumber
\end{eqnarray}
where the correlator $\left\langle \Delta \Omega_{t_1 z}\Delta\Omega_{t_1 'z}\right\rangle$ is given by Eq. (29). According to Fig. 7, around the symmetric point $\varphi =0$ the $+$ mode is only essential in Eq. (33) and we use $\left\langle r_t \right\rangle =r_0\exp [(i\Delta\Omega_+ -\nu_+ )t]$. Within the logarithmic approach, $\left\langle\delta N_{\ldots}\delta N_{\ldots}\right\rangle$ is replaced by the constant [see Eq. (44)] and after the straightforward integrations the mean-square fluctuation of amplitude takes form:
%48
\begin{eqnarray}
\left\langle |\delta r_t |^2 \right\rangle  \approx |r_0 |^2 \left[\left( {\frac{{\nu _{\varphi  = 0} }}{{\Delta \Omega _ +  }}} \right)^2 \frac{{\Lambda _2 }}{\pi } + 2\left( {\frac{\widetilde \nu _1}{\Delta \Omega _ +}} \right)^2  \left( {\frac{\Lambda _1}{\pi }} \right)^2\right]  \nonumber    \\
\times \left( {1 + e^{ - 2\nu_+ t} \normalsize-2e^{-2\nu_+ t} \cos \Delta\Omega_+ t} \right) .  ~~~~~~~~
\end{eqnarray}
Here the first and second term in $[\ldots ]$ are due to the long- and short-scale contributions to fluctuations [see similar terms in Eq. (27)] and this factor is estimated as logarithmically weak, $[\ldots ]\sim 1/\Lambda_{1,2}<1$, see Eqs. (24), (32), and (34).

At $\varphi =0$ the requirement for averaged description of decoherence is $\sqrt{\left\langle |\delta r_t |^2 \right\rangle}<r_0 e^{-\nu_+ t}$. In analogy to the previous section, the results of Sect. IV B are valid if $\nu_+ t<\ln (1/[\ldots])$ and, for a typical parameters, fluctuations are weak over the time interval $\nu_+ t<3$. A similar condition remains valid over the dip region $|\varphi |<0.3\varphi_1$ (see Fig.8) where both ${\cal X}_{\varphi}$ and ${\cal Z}_{\varphi}$ are changed only due to numerical factors $\sim 1$. Thus,the fluctuations contribution is suppressed only in the saturation region, $|\varphi |\geq\varphi_1$ while the averaged description over the dip is valid during a short enough interval.

%%%%%%%%%%%%%%%%%%%%%%%%%%%%%%%%%%%%
\section{Concluding remarks }
We present a comprehensive investigation of the dissipative dynamics of a flux qubit, describing interwell and interlevel relaxation as well as decoherence caused by the 1/f noises pass through the SQUID and the LC-contour. We analyze the rates of these processes and the renormalization of gap frequency versus the tilt and control fluxes taking into account correlations between the noises; this permits one to characterize the qubit-noise interaction. We show how rare fluctuations limit the averaged description at tails of relaxation. Under typical level of noises, the results obtained give contributions comparable to the recent experimental data on the interlevel population relaxation and the interwell tunneling.

Our consideration is based on a several assumptions which are shortly discussed below. 
\begin{itemize}
\item[(a)]\vspace{-0.25 cm} Description of the flux qubit is based on the effective circuit formed by the effective Josephson junction shunted by the LC-contour, instead of the SQUID loop, which is shunted by the transmission line. It is a good approach for the low-frequency region, far below the characteristic frequency of the LC-contour which is $>$10 GHz for a device with typical parameters.
\item[(b)]\vspace{-0.25 cm} Consideration of 1/f noise as a classical random flux is valid for frequencies below 1$\div$2 GHz where the interaction with high-frequency bosons can be omitted \cite{15}. As a result, qubit approachs to the equi-populated distribution, $\Delta n_t |_{\nu_1 t\gg 1}\to 0$, while the equilibrium distribution can be reached due to the qubit-boson interaction, during times beyond the scales considered here. 
\item[(c)]\vspace{-0.25 cm} Because of the logarithmic character of the cut-off, one can use the abrupt cut-off at $\omega \sim\omega_m$ in Eq. (3) and below. A possible deviation from 1/f spectrum, e.g. due to the size effect in the transmission line \cite{22}, may be described similarly after choosing a specific parameters of device.  
\item[(d)]\vspace{-0.25 cm} The qubit-noise interaction is studied by adding random fluxes to the tilt and control fluxes and by taking into account correlations between these sources. The coupling levels in $tl$-, $j$-, and $c$-channels are given by the phenomenological parameters $\alpha_{tl,j,c}$. Both a microscopic study of a 1/f noise mechanism and a detail description of a noise effect on the SQUID loop and the LC transmission line are beyond of the scope of the paper and can be performed for specific devices. 
\item[(e)]\vspace{-0.25 cm} The idealized initial conditions were used at $t=0$ without any discussion of a temporal evolution at $t<0$. A more detailed description requires an analysis of the protocol of resonant tunneling \cite{14,15} and the mechanisms of the ultrafast interlevel excitation, see \cite{9,25} and references therein. Here we neglect a possible uncertainties during the initialization and readout stages. 
\item[(f)]\vspace{-0.25 cm} We restrict ourselves by the weak-fluctuation regime, when the averaged dissipative dynamics describes an evolution of qubit. Level of fluctuations gives the limitations of the averaged description at tails of relaxation. In principle, a chaotic regime contains an additional information on the qubit-environment interaction but an analysis of the two-point spectral functions requires a special consideration.
\end{itemize}\vspace{-0.25 cm}

Now we discuss the current experimental data for the LC-shunted qubits and possibilities for verification of the results obtained. In spite of these qubits were employed for demonstration of the multi-qubit clusters \cite{1,2,26}, a complete spectroscopic characterization of this type of qubits is not available \cite{23}. Recent measurements of the population relaxation rate in the region above 0.5 GHz \cite{15} give the amplitude and half-width of peak in agreement with the results of Sect. IV B but the 1/f spectral dependency is modified due to the soft barrier effect. The decoherence processes were not analyzed in \cite{15,23} but such a data are necessary in order to verify of the $j$- and $c$-channels contributions which determine the depth and asymmetry of the dip. A study of the other peculiarities discussed in Sect. IV C (the two-mode evolution and the renormalization of gap) may be complicated due to the fluctuation-induced  restrictions on the averaged response. The incoherent tunneling rate reported in \cite{14,23} was in agreement with the model \cite{21} when the levels are modulated by coupling to the boson thermostat. It lead to the temperature dependent asymmetry of the tunneling peak. Recent measurements \cite{15} show an additional temperature-independent asymmetry  which can be caused by the $c$-channel contribution, see Sect. III. In order to confirm this mechanism, one needs to demonstrate a changing of the sign of asymmetry for the parallel and antiparallel directions of $\varphi$ and $\psi$, see Fig. 1(a). Thus, the above-discussed partial experimental data does not permit to characterize the qubit-noise interaction completely. It is necessary to perform all possible measurements for the same device and, using an appropriate model of the device, find parameters which will describe all data. Similar program can be develop for other types of qubits, with  an effective circuits different from Fig. 1(b), such as the capacitively shunted flux qubit \cite{10} or another variants of transmon \cite{12,27}.  

To conclude, the obtained results convincingly demonstrate that the transient dynamics of the flux qubit should be analyzed beyond of the simplified two-level model which includes only a noise-induced modulation of the interlevel gap. Our consideration, which is based on the lumped-element approach with detailed description of noise effects, opens the way to characterize the flux qubit interacting with low frequency noise and to enhance a fidelity of the device. We believe that a similar study of another types of qubits, the interqubit connections, and the multi-qubit clusters will improve parameters of the quantum hardware.
%%%%%%%%%%%%%%%%%%%%%%%%%%%%%%%%%%%%%%%
\appendix*    
\section{Averaged kernels} 
Here we consider averaging over the random Gaussian noises in $tl$-, $j$-, and $c$-channels for the kernels employed in sections III-V. We  begin with the kernel $K_{\Delta t}$ in the averaged integro-differential equation (11). Explicit expression for $K_{\Delta t}$  is obtained below with the use of $\omega_{tx}$ and $\omega_{tz}$, given by Eq. (5), and the phase factor $\theta _{tt'}  = \theta _{\Delta t}  + \Delta\theta_{tt'}$:
%A1
\begin{eqnarray}
K_{\Delta t}\!  =\! \cos \theta _{\Delta t}\!\! \left[ {\omega _{10}^2 \left\langle {\cos \Delta \theta _{tt'} } \right\rangle\!  +\! \omega _j^2 \left\langle {\Delta \psi _t \Delta \psi _{t'} \cos \Delta \theta _{tt'} } \right\rangle } \right]   \nonumber  \\
- \omega _j \omega _{10} \sin \theta _{\Delta t} \left\langle {(\Delta \psi _t  + \Delta \psi _{t'} )\sin \Delta \theta _{tt'} } \right\rangle ~, ~~~~~~~~
\end{eqnarray}
where the unperturbed phase and the noise contribution to phase are given by $\theta_{\Delta t}=2\widetilde \omega \varphi \Delta t$ and $\Delta \theta_{tt'}=2\widetilde \omega\int_{t'}^td\tau\Delta \varphi_\tau$, respectively. To carry out the averaging of $\propto\omega _{10}^2$ contribution, one should consider all possible pairings in the  expansion of cosine and the total number of such pairings in the term of the order $2n$ is equal to $(2n)!/2^n n!$, where factor $2^n$ is due to symmetry of correlator with respect to $t\leftrightarrow t'$ and factor $n!$ gives number of permutations of $n$ pairs. The infinite sum over $n$ is transformed to an exponent (compare the similar averaging over spatial domain \cite{28}) so that
%A2
\begin{equation}
 \left\langle{\cos\Delta\theta_{tt'}} \right\rangle\! =\! \exp\! \left\{\! -2\widetilde \omega^2 \!\!\int_{t'}^t\!\! {d\tau } \!\!\int_{t'}^t\!\! {d\tau '} \left\langle {\Delta \varphi_\tau \Delta \varphi _{\tau '} } \right\rangle \! \right\}\!\! \equiv\! e^{-\Gamma_{\Delta t}} .
\end{equation}
The correlator under integrals over $\tau$ and $\tau '$ is defined by Eq. (3) so that $\Gamma_{\Delta t}$ is determined by Eq. (13a).

After the similar expansion of cosine, the proportional to $\omega_j^2$ contribution in Eq. (A1) takes form 
%A3
\begin{eqnarray}
 \left\langle{\Delta\psi_t \Delta\psi_{t'}\cos\Delta\theta_{tt'}} \right\rangle\!\! = e^{-\Gamma_{\Delta t}}\left[ \left\langle {\Delta \psi_t \Delta \psi _{t'} } \right\rangle \right.  ~~~   \\
\left. - (2\widetilde\omega )^2\! \int_{t'}^t\! {d\tau } \left\langle {\Delta \psi _t \Delta \varphi _\tau  } \right\rangle\! \int_{t'}^t\! {d\tau '} \left\langle {\Delta \psi _{t'} \Delta \varphi _{\tau '} } \right\rangle  \right]  ,  \nonumber  
\end{eqnarray}
and it involves both $\propto\alpha_j$ and $\propto\alpha_c$ addendums. The last addendum contains the same integrals which are dependent on $\Delta t$ and are transformed into Eq. (13b). Similar averaging for $\propto\omega_j\omega_{10}$ contribution in Eq. (A1) is performed after the expansion of the sine:
%A4
\begin{eqnarray}
 \left\langle{(\Delta\psi_t  + \Delta\psi_{t'})\sin\Delta\theta_{tt'}}\right\rangle\!\! =2\widetilde\omega e^{-\Gamma_{\Delta t}}   \\
\times \int_{t'}^t\!\! {d\tau}\left\langle (\Delta\psi_t  + \Delta\psi_{t'})\Delta \varphi _{\tau}  \right\rangle  ~~~~~~   \nonumber
\end{eqnarray}
and this contribution is written through ${\cal B}_{\Delta t}\propto W_{\Delta t}$, see Eq. (12). 

For the weak-coupling regime, the averaged kernel of Eq. (21) is given by 
%A5
\begin{equation}
N_{\Delta t}  = \cos (\Omega _\varphi  \Delta t)\left\langle {\Omega _{tx} \Omega _{t'x} \cos \left( {\int_{t'}^t {d\tau } \Delta \Omega _{\tau z} } \right)} \right\rangle  
\end{equation}
and with $\propto\alpha_{\ldots}$-accuracy it can be written as $N_{\Delta t}\approx \cos (\Omega_\varphi \Delta t)\left\langle \Omega _{tx} \Omega _{t'x}\right\rangle$ with the use $\cos\left({\int_{t'}^t\ldots} \right)\approx 1$. This approximation gives $\propto{\cal X}_\varphi$ contributions in Sects. IV A and IV B.

Under examination of the fluctuation effect (Sect. V) one needs to calculate the four-point correlators of the random sources. The correlator of fluctuations in Eq. (38) is written through the correlator $\left\langle\cos\theta_{t_1 t'_1 }\cos\theta_{t_2 t'_2}\right\rangle$ which is transformed through $\left\langle\cos (\theta_{t_1 t'_1 }\pm\theta_{t_2 t'_2})\right\rangle$ and using Eq. (A2) one obtains 
%A6
\begin{eqnarray}
\left\langle\cos\theta_{t_1 t'_1 }\cos\theta_{t_2 t'_2}\right\rangle = \cos \theta _{\Delta t_{12} } \cos \theta _{\Delta t'_{12} } e^{ - \Gamma _{\Delta t_{12} }  - \Gamma _{\Delta t'_{12} } }   \nonumber   \\
\times\exp \left[ {\alpha _{tl} (2\tilde \omega )^2 \int_{t_2 }^{t_1 } {d\tau } \int_{t'_2 }^{t'_1 } {d\tau } 'w_{\omega _m |\tau  - \tau '|} } \right]  . ~~~~~~~~~
\end{eqnarray}
The correlator (39) is written after expansion of $\exp [\alpha_{tr}\ldots ]$ with respect to $\propto\alpha_{tr}$ contributions. For the weak-coupling regime, the four-point fluctuations of kernels in Eqs. (43) and (47) are transformed as follows:
%A7
\begin{eqnarray}
 \left\langle\delta N_{t_1 t_2}\delta N_{t'_1 t'_2}\right\rangle= \alpha _{tl} (2\widetilde \omega )^4 \left( w_{\omega _m |t_1  - t'_1 |} w_{\omega _m |t_2  - t'_2 |} \right. \nonumber \\
\left. + w_{\omega _m |t_1  - t'_2 |} w_{\omega _m |t_2  - t'_1 |}  \right) . ~~~~~~
\end{eqnarray}
Within the logarithmic approach, this correlator gives the time-independent factor used in Eqs. (44) and (48).

\end{document}